\documentclass{aiaa-tc}

\usepackage{wrapfig}
\usepackage{threeparttable}
\usepackage{dcolumn}
\newcolumntype{d}{D{.}{.}{-1}}
\usepackage{nomencl}
\makeglossary
\usepackage{subcaption}
\usepackage{caption}
\usepackage{wrapfig}
\usepackage{fancyvrb}
\fvset{fontsize=\footnotesize,xleftmargin=2em}
\usepackage{lettrine}

\usepackage{booktabs}
\usepackage{bm, amssymb, amsmath, array, pdfpages}
\usepackage{amsmath,amssymb,color,graphicx,pdfsync,overpic,color,epstopdf,rotating,dashrule,float,bm}
\usepackage{float}

\usepackage{setspace}
\usepackage{comment}
\usepackage{changepage}

\usepackage{titlesec}

\newcommand{\diff}[2]{\frac{\partial #1}{\partial #2}}
\newcommand{\beq}[0]{\begin{equation}}
\newcommand{\eeq}[0]{\end{equation}}
\newcommand{\beqq}[0]{\begin{equation*}}
\newcommand{\eeqq}[0]{\end{equation*}}
\newcommand{\bs}[1]{\boldsymbol{#1}}
\newcommand{\bv}[1]{\mathbf{#1}}
\DeclareMathOperator{\var}{Var}

\graphicspath{{../Figures/}}

\title{Uncertainty Quantification for Cargo Hold Fires}

\author{%
  Anthony M. DeGennaro\thanks{Department of Mechanical and Aerospace
    Engineering; Student Member, AIAA.}
  \ ,
  Mark W. Lohry\thanksibid{1}
  \ ,
  Luigi Martinelli\thanks{Department of Mechanical and Aerospace Engineering;
    Associate Fellow, AIAA.}
  \ ,
  Clarence W. Rowley\thanksibid{2}\\
  {\normalsize\itshape
   Princeton University, Princeton, NJ, 08540, USA}\\
 }

 \AIAApapernumber{200?-????}
 \AIAAconference{Conference Name, Date, and Location}
 \AIAAcopyright{\AIAAcopyrightD{200?}}

\def\ip<#1,#2>{\left\langle #1,#2\right\rangle}

\begin{document}

\maketitle

\begin{abstract}
  The purpose of this study is twofold -- first, to introduce the
  application of high-order discontinuous Galerkin methods to
  buoyancy-driven cargo hold fire simulations; second, to explore
  statistical variation in the fluid dynamics of a cargo hold fire
  given parameterized uncertainty in the fire source location and
  temperature. Cargo hold fires represent a class of problems that
  require highly-accurate computational methods to simulate
  faithfully. Hence, we use an in-house discontinuous Galerkin code to
  treat these flows. Cargo hold fires also exhibit a large amount of
  uncertainty with respect to the boundary conditions. Thus, the
  second aim of this paper is to quantify the resulting uncertainty in
  the flow, using tools from the uncertainty quantification community
  to ensure that our efforts require a minimal number of
  simulations. We expect that the results of this study will provide
  statistical insight into the effects of fire location and
  temperature on cargo fires, and also assist in the optimization of
  fire detection system placement.

\end{abstract}

\section{Introduction}

Federal aviation regulations require that all large passenger aircraft
have fire detection and suppression systems in all cargo
compartments. Several different detection methods are generally used
together, such as sensors for temperature, carbon monoxide, smoke
particulate, radiation, and optical detection. These sensors are
required to detect the fire within 60 seconds of fire ignition. Certification
of these systems currently requires expensive ground and in-flight
testing. Current fire detection certification focuses on experiments
using a small fire in empty cargo holds, such as the narrow-body
Boeing 707 fuselage located at the Federal Aviation Administration
William J. Hughes Technical center in Atlantic City, New
Jersey\cite{Oztekin}. Simulating a single fire case is a well-posed
problem and relatively straightforward, but of limited utility. Due to
the costs associated with these types of experiments, testing a wide
variety of fire sources, positions, and compartment cargo cluttering
is not feasible.

CFD tools that can accurately simulate heat and particulate transfer
in fire-induced flow in cargo holds can potentially reduce these
certification costs by reducing the amount of experimental work
necessary. Simulations can then be used to assess the effectiveness of
a particular detector placement, as well as optimize their placement
in a given cargo hold. The allure of CFD tools is the reduction of
monetary costs associated with certification tests; however, a
drawback is the associated computational expense. In light of this, an
issue that needs to be addressed is how to accurately quantify the
uncertainty associated with randomly variable boundary conditions
(eg., fire source location or temperature) while using the least
amount of CFD simulations possible.

This work has two main objectives. The first is to establish efficient
and accurate CFD tools that can be used to simulate cargo fires over a wide range
of parameters. For these simulations we develop an in-house high-order
accurate discontinuous Galerkin (DG)\cite{hesthaven2008nodal} flow
solver on unstructured meshes. The DG scheme approach is well-suited
for computing the turbulent, vorticity-dominated buoyancy-driven flows
observed in cargo hold, and unstructured meshes allows one to compute
on a complex domain such as those encountered in cluttered cargo
holds.

The second objective is to apply techniques of uncertainty
quantification to explore the statistical effects of parameterized
boundary condition uncertainty with the ultimate goal of optimizing
the placement of fire detection systems. In particular, we will be
using Polynomial Chaos Expansions (PCE) to achieve this, as this
method is efficient and accurate. In order to assess the feasibility
of these methods to the problem at hand, we are restricting the
problem to the 2-dimensional cross-section of the cargo hold.

\section{Simulation methodology}

\subsection{Discontinuous Galerkin simulation tool}

It is well known that traditional low-order $O(\Delta x^2)$ flow solvers are excessively dissipative for vorticity-dominated flows such as those seen in fires. Adequate resolution of vorticity convection far from its generation source typically requires either a prohibitively fine mesh or a higher-order representation of the flow solution. The in-house simulation tool used in this work is a nodal discontinuous Galerkin (DG) flow solver for the compressible Navier-Stokes equations with buoyancy effects, discretized with an unstructured mesh suitable for complex geometries and arbitrarily-high order of accuracy. The spatial discretization used here follows that detailed by Hesthaven and Warburton\cite{hesthaven2008nodal}, and is briefly summarized here.

For a multi-dimensional conservation law of quantity $u$, flux $\bv{f}$, and source $\Psi$
 \beq
 \diff{u(\bv{x},t)}{t} + \nabla \cdot \bv{f}(u(\bv{x},t),\bv{x},t) = \Psi(\bv{x},t)
 \label{eq:conslaw}
 \eeq
 \noindent
 the quantities can be approximated by an expansion
 \beq
 u (\bv{x},t) \approx u_h(\bv{x},t) = \sum_{i=1}^{N_p} u_h(\bv{x}_i,t) \bs{l}_i(\bv{x})
 \eeq
 \noindent
 where $\bs{l}_i(\bv{x})$ is the multidimensional Lagrange polynomial defined by grid points $\bv{x}_i$, and $N_p$ is the number of nodes in the element, $N_p = (N+1)(N+2)/2$ for a triangular element of polynomial order $N$.

 Taking the product of this with the same Lagrange polynomial $\bs{l}_j$ serving as a test function and integrating by parts on the spatial component over an element $V$ with surface $S$ yields
\beq
\int_V \left( \diff{u_h}{t} \bs{l}_j(\bv{x}) -\bv{f}_h \cdot \nabla \bs{l}_j(\bv{x}) - \Psi_h \bs{l}_j \right)\ dV =  -\int_S \bv{f}^\star \bs{l}_j(\bv{x}) \cdot \bv{n} \ dS 
\label{eq:weakform}
\eeq
\noindent
where flux $\bv{f}^\star$ is the numerical flux, uniquely defined at element interfaces. In this work the inviscid components of flux are computed using the local Lax-Friedrichs flux splitting, and the viscous flux components use a centered average.

Time integration is performed using the implicit 3rd order backward difference formula
\beq
\frac{du}{dt} \approx \left(u^{n+1} - \frac{18}{11}u^{n} + \frac{9}{11}u^{n-1} - \frac{2}{11}u^{n-1}\right) / \left( \frac{6}{11} \Delta t \right)
\eeq
where $\Delta t$ is the discrete time step size and $n$ the time step index.

This discretization leads to a non-linear system of algebraic equations to be solved at each time step. The non-linear system can be written as $\bv{F}(\bv{u}) = 0$, and Newton's method can be used with the iterative step index $k$,
\beq
\bv{F}(\bv{u^{k+1}}) = \bv{F}(\bv{u^{k}}) + \bv{F'}(\bv{u^{k}}) (\bv{u^{k+1}}- \bv{u^{k}}) 
\eeq
resulting in a sequence of linear systems
\beq
\bv{J}(\bv{u^{k}}) \delta \bv{u^k} = -\bv{F}(\bv{u^{k}}),\quad \bv{u^{k+1}}=\bv{u^{k}}+\delta \bv{u^k}
\eeq
for the Jacobian $\bv{J} = \bv{F'}(\bv{u})$. The Jacobian matrix $\bv{J}$ is a very large sparse matrix which can be prohibitively expensive to store in computer memory. Fortunately the Krylov subspace methods for the solution of linear algebraic systems do not require this matrix itself, but only the matrix-vector product. This can be approximated by a finite difference
\beq
\bv{J} \delta \bv{u} \approx [\bv{F}(\bv{u} + \epsilon \delta \bv{u}) - \bv{F}(\bv{u})]/\epsilon
\eeq
for a small ($\sim 10^{-6}$) parameter $\epsilon$. In this work the restarted GMRES algorithm is used for the solution of the linear systems at each Newton iteration, with the Newton method progressing until a desired convergence tolerance is reached and the physical time step is advanced. This approach for solving non-linear systems by coupling matrix-free Krylov iterative methods for linear systems with Newton iterations is known as a ``Jacobian-free Newton-Krylov'' (JFNK) method, and is detailed in the review paper by Knoll \& Keyes\cite{Knoll2004}.

\subsection{Cargo hold geometry and boundary conditions}

\begin{wrapfigure}{r}{0.5\textwidth}
  \centering
  \begin{subfigure}[b]{0.48\textwidth}
    \includegraphics[width=\textwidth,trim = 5mm 5mm 5mm 5mm,clip]{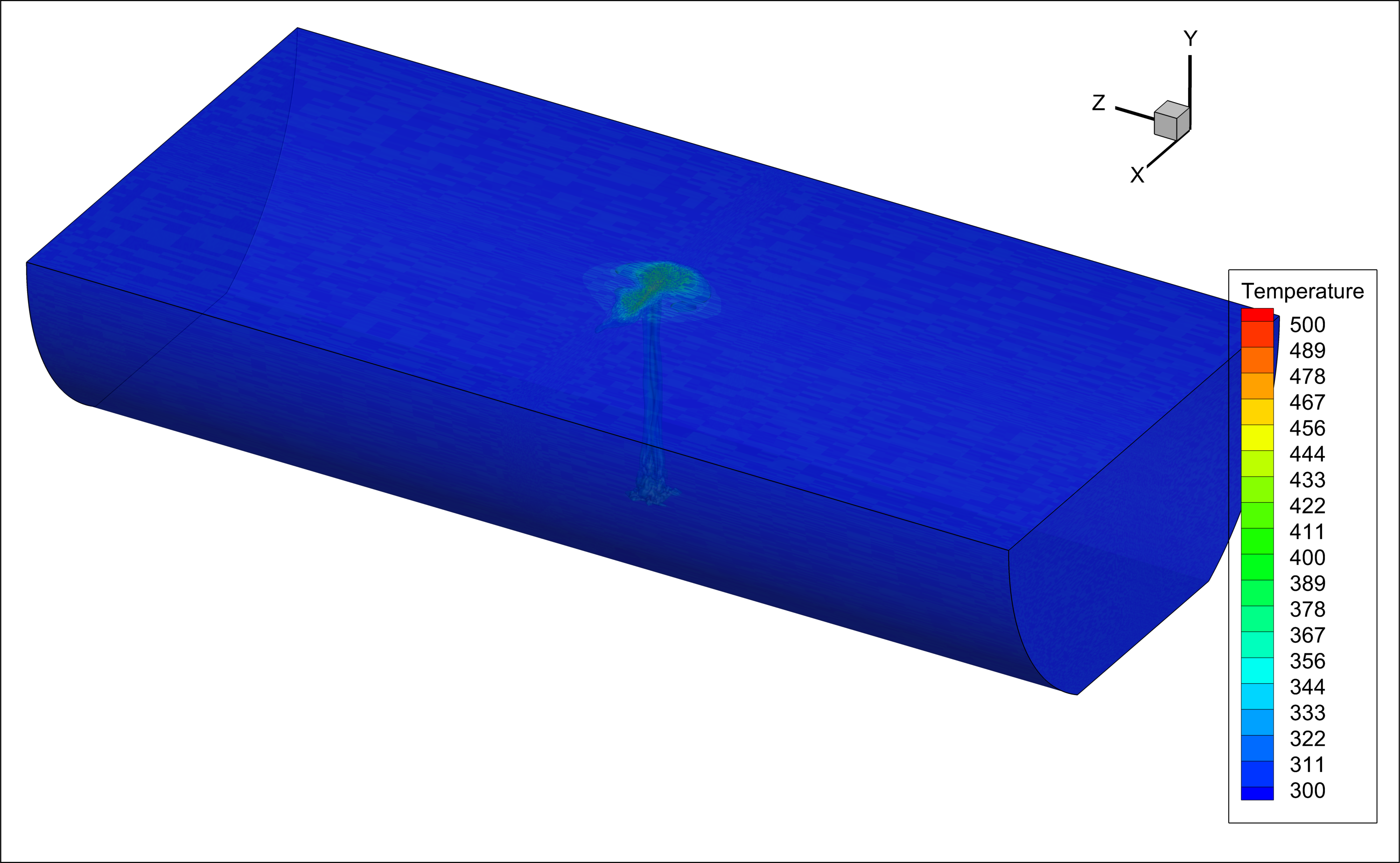}
    \caption{Temperature field after start-up of a small fire in the center.}
    \label{fig:geom3d}
  \end{subfigure} \begin{subfigure}[b]{0.48\textwidth}
    \includegraphics[width=\textwidth,trim = 5mm 30mm 5mm 5mm,clip]{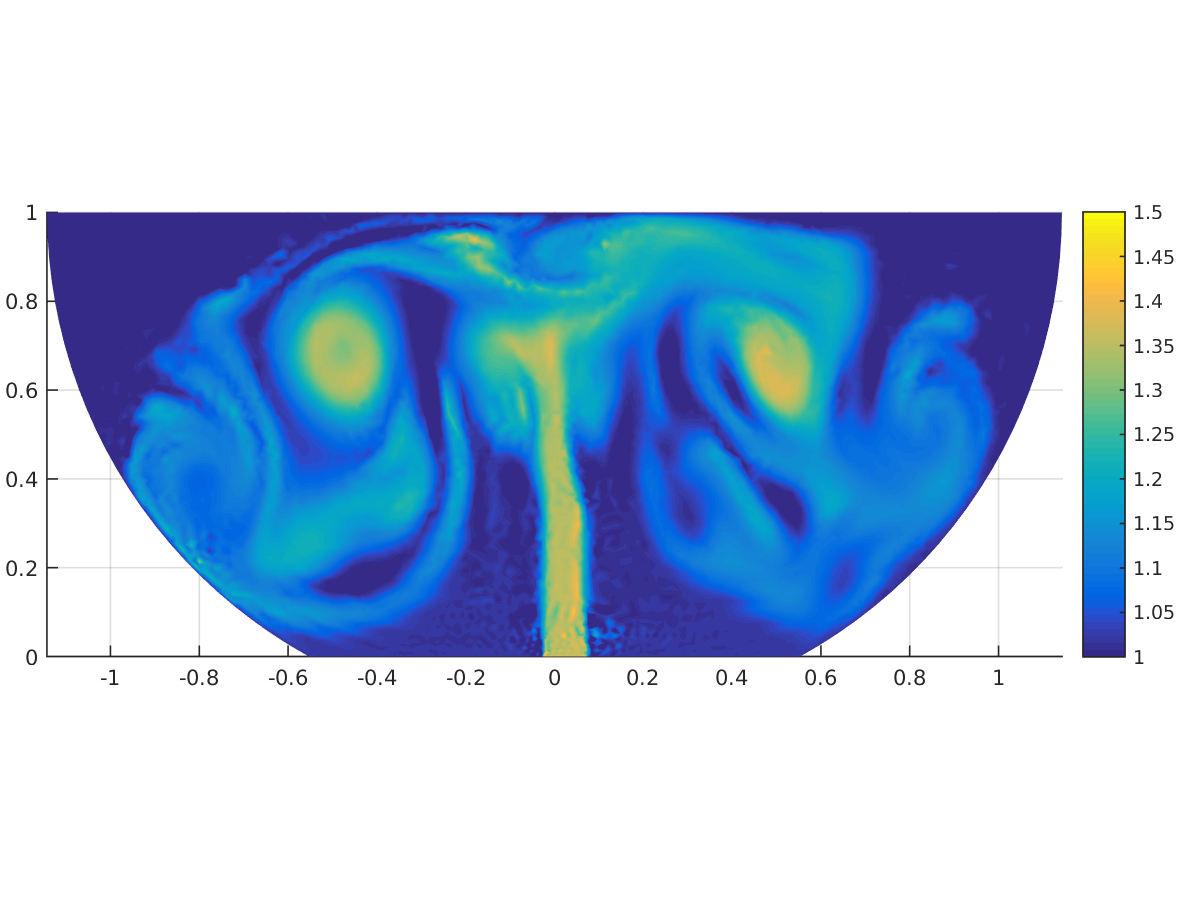}
    \caption{Flow driven by a heat source in a 2D cross-section. Colormap shown is temperature normalized by the initial bulk temperature.}
    \label{fig:cleanstill}
  \end{subfigure}
  \caption{Example flowfields of buoyancy-driven flow in Boeing 707 cargo hold geometry.}
\end{wrapfigure}

The geometry of interest here is the forward cargo compartment of a
Boeing 707. This geometry shortly after a small fire is started in the
center can be seen in figure~\ref{fig:geom3d}. The flow is entirely
driven by buoyant effects due to the local heating produced by the
fire. We note that a short distance away from the fire source, there
is no longer a significant effect on the dynamics of the flow due to the actual
chemical combustion process taking place. This type of flow can therefore be
accurately modeled as a heat source addition into
non-reactive air, freeing us from the need to tackle the
computationally expensive details of the combustion problem. Experimental results of the full 3D case and background on this problem can be found in work by Oztekin et al\cite{Oztekin,Oztekin2012}.

A typical simulation of a 2D cross-section of the cargo geometry
(computed using our in-house DG code) can be seen in
figure~\ref{fig:cleanstill}. A turbulent plume rising from the heat
sources drives vortical flow around the compartment feeding back into
itself at the bottom. We note recirculation regions in both upper
corners leading to stagnation regions where streamlines are
separating, indicating a sensor position there would be less effective
than at other locations. The turbulent, buoyant flow is
instantaneously asymmetric, but statistically averaged is largely
symmetric due to symmetric boundary conditions. The base of the
geometry is $1.107m$ wide, and the ceiling is $2.286m$ wide and $1m$
tall.

In this work, we restrict the analysis to a 2D cross-section of a
cargo hold. All boundary conditions are isothermal, with the majority
of the wall boundary fixed to the initial bulk temperature
non-dimensionalized to $T_{\infty}=1$. A $0.1m$ wide section of the
floor is then set to an isothermal condition at a multiple of the bulk
temperature in order to model a heat source. The temperature source
$T_s$ is examined in the range between $T_s = 1.2$ and $1.5$, and the
temperature location $x_s$ in the range between $x_s = 0.0$
(centerline) and $0.503m$ (the rightmost possible location for $0.1m$
wide source.) Due to symmetry, sources need only be placed to one side
of the geometry in order to analyze sources at a reflected point along
the floor.

All DNS simulations here are performed using cubic ($N=3$) elements, with
the 2D meshes consisting of approximately $1500$ triangular
cells. This results in $10$ nodes per cell for each of the $4$
quantities (density, $x$ and $y$ momentum, and energy) to be solved,
for a total of $\sim 60,000$ degress of freedom.

\begin{figure}
  \centering
  \begin{subfigure}[b]{0.32\textwidth}
    \includegraphics[width=\textwidth,trim = 0mm 35mm 0mm 35mm,clip]{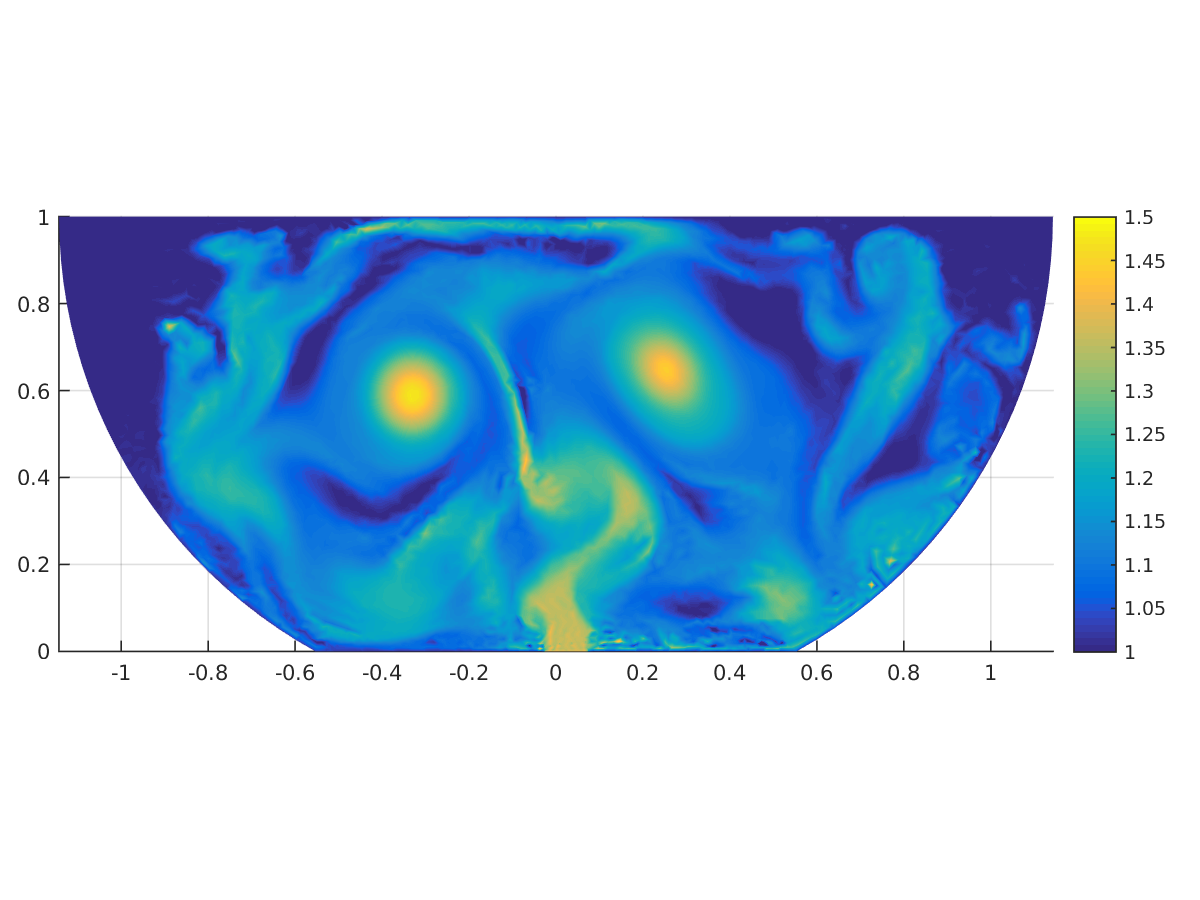}
    \caption{$x_s = 0.024m$.}
  \end{subfigure} \begin{subfigure}[b]{0.32\textwidth}
    \includegraphics[width=\textwidth,trim = 0mm 35mm 0mm 35mm,clip]{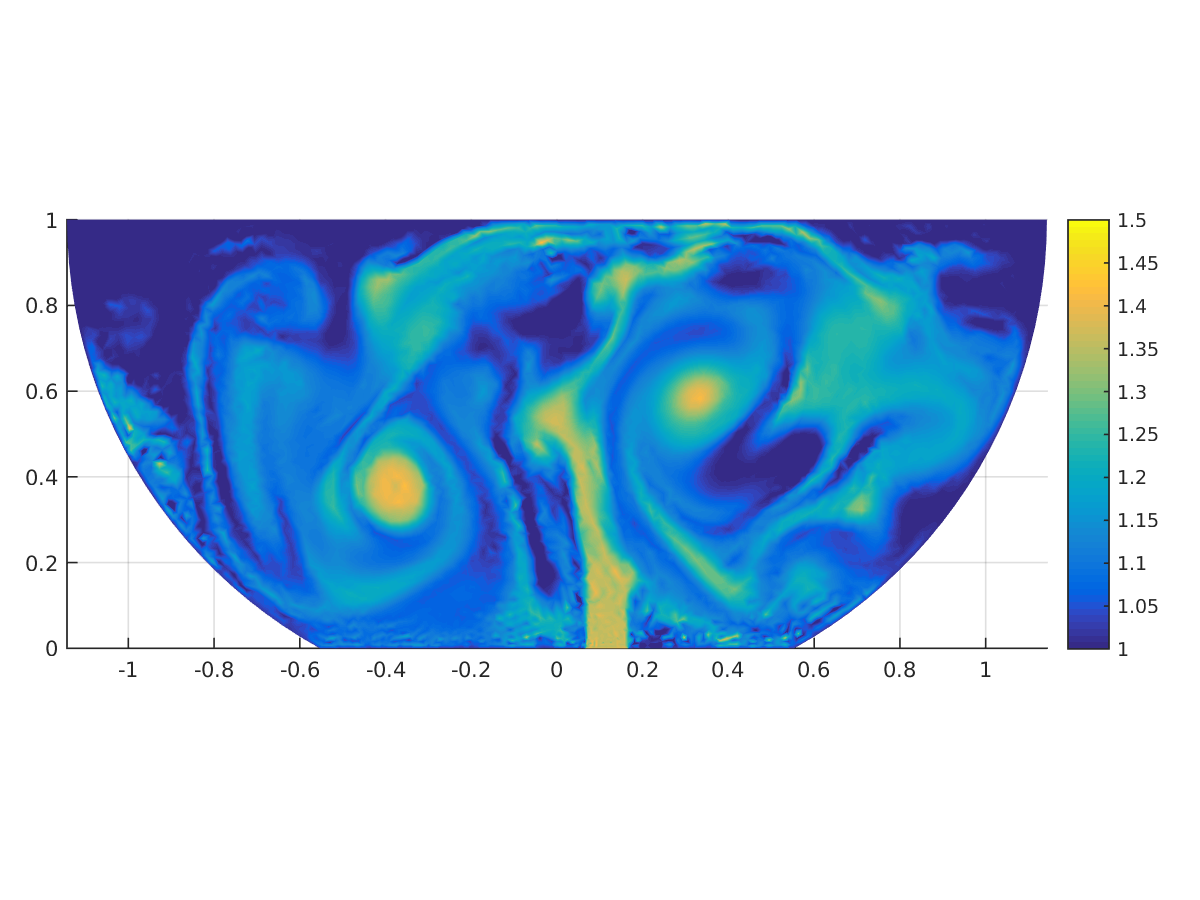}
    \caption{$x_s = 0.116m$.}
  \end{subfigure} \begin{subfigure}[b]{0.32\textwidth}
    \includegraphics[width=\textwidth,trim = 0mm 35mm 0mm 25mm,clip]{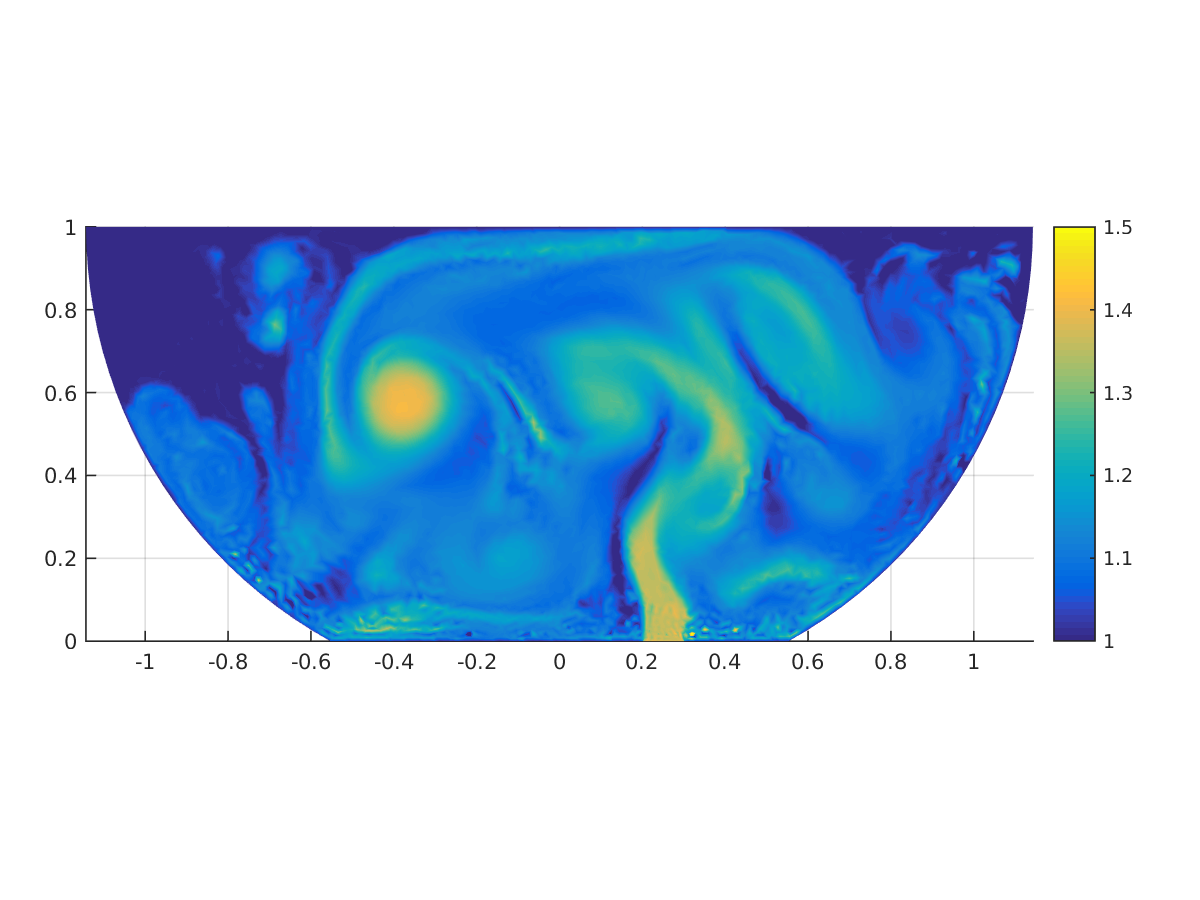}
    \caption{$x_s = 0.262m$.}
  \end{subfigure} \begin{subfigure}[b]{0.32\textwidth}
    \includegraphics[width=\textwidth,trim = 0mm 35mm 0mm 35mm,clip]{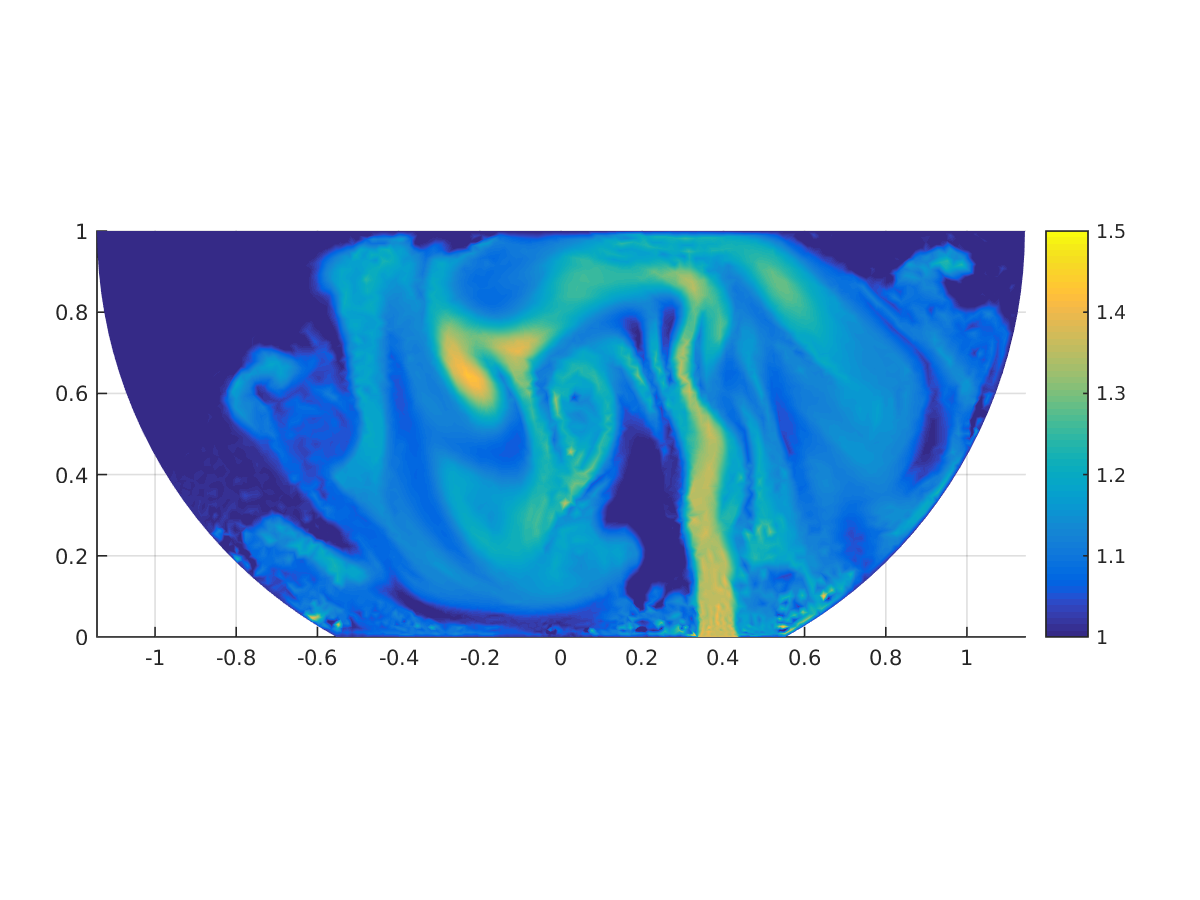}
    \caption{$x_s = 0.387m$.}
  \end{subfigure} \begin{subfigure}[b]{0.32\textwidth}
    \includegraphics[width=\textwidth,trim = 0mm 35mm 0mm 35mm,clip]{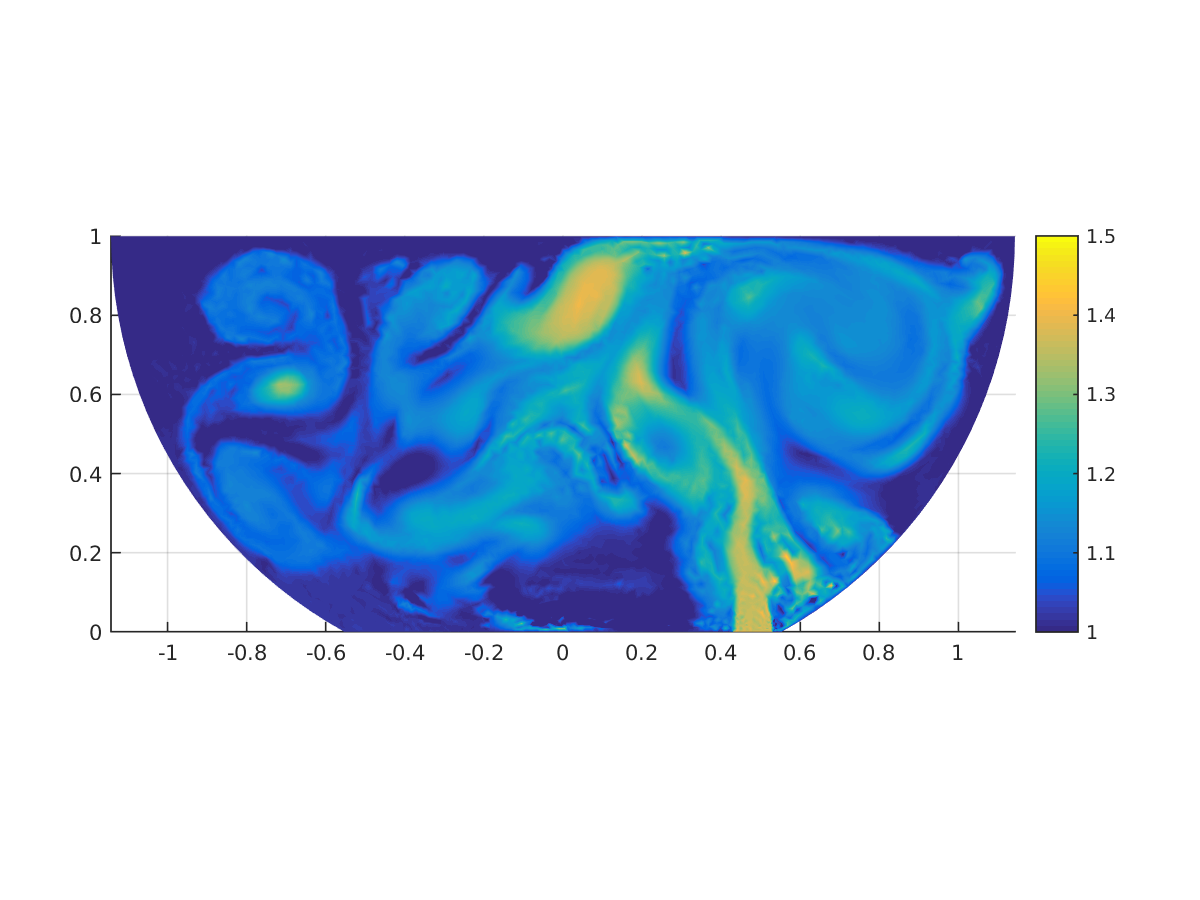}
    \caption{$x_s = 0.480m$.}
  \end{subfigure}
  \caption{Temperature fields for $T_s=1.486$ source at the 5 source locations, time $t=10s$ after startup.}
  \label{fig:fixedheat}
\end{figure}

\begin{figure}
  \centering
  \begin{subfigure}[b]{0.32\textwidth}
    \includegraphics[width=\textwidth,trim = 0mm 35mm 0mm 35mm,clip]{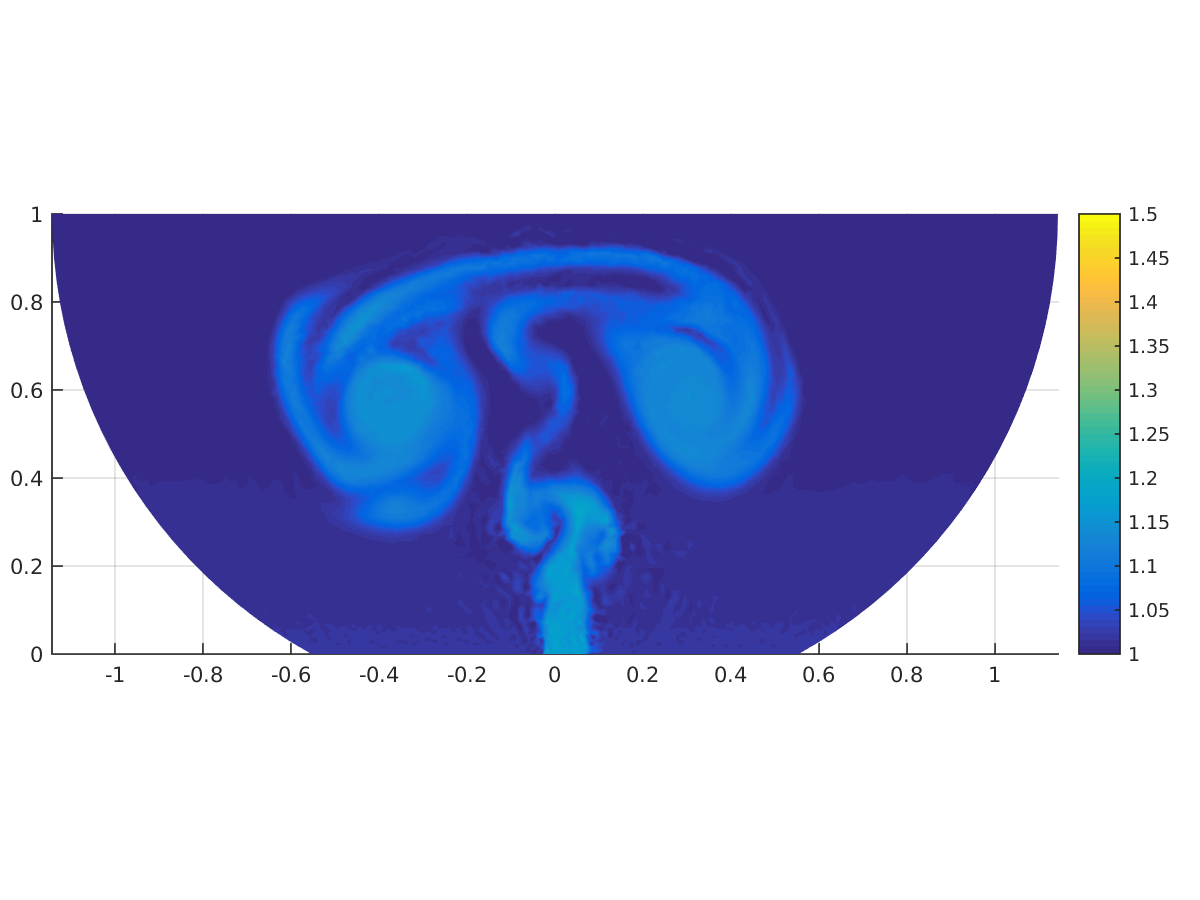}
    \caption{$T_s = 1.214$.}
  \end{subfigure} \begin{subfigure}[b]{0.32\textwidth}
    \includegraphics[width=\textwidth,trim = 0mm 35mm 0mm 35mm,clip]{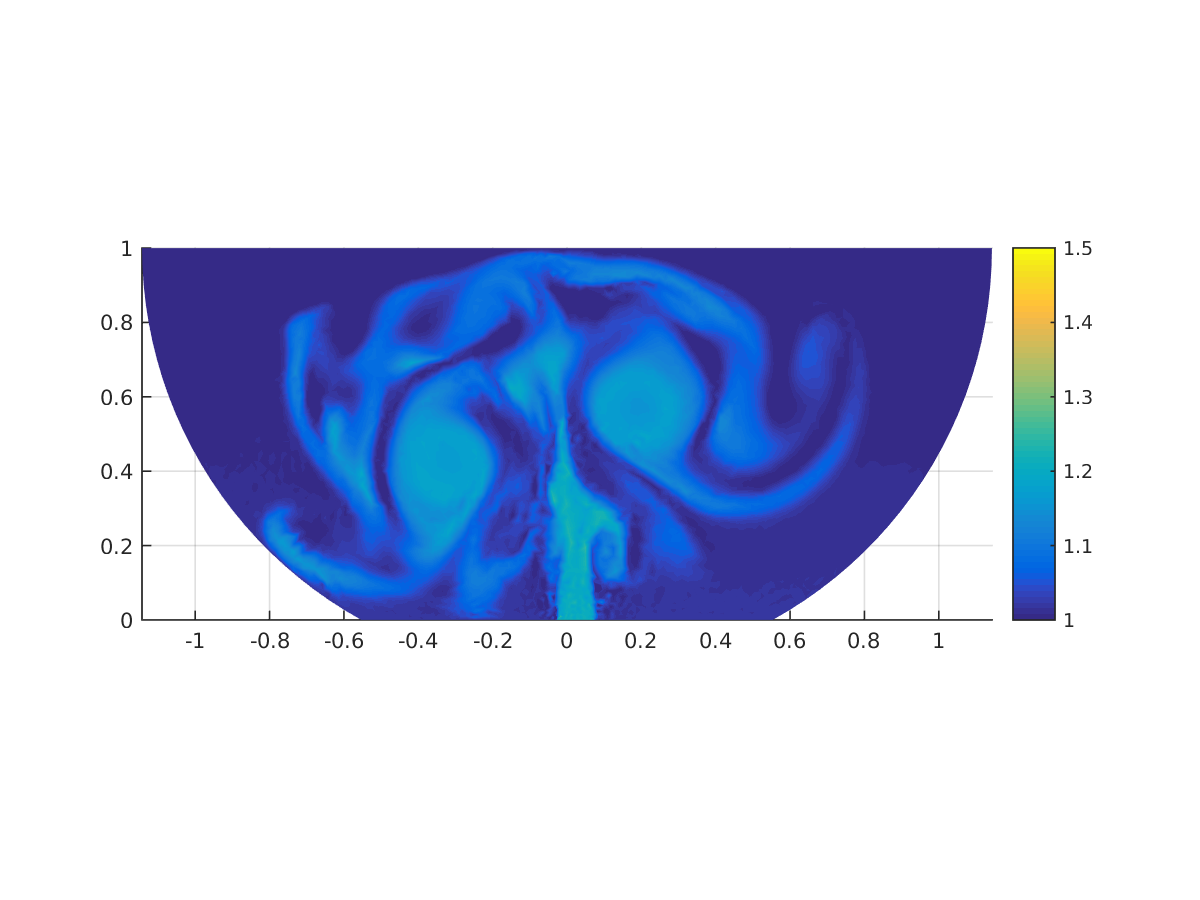}
    \caption{$T_s = 1.269$.}
  \end{subfigure} \begin{subfigure}[b]{0.32\textwidth}
    \includegraphics[width=\textwidth,trim = 0mm 35mm 0mm 25mm,clip]{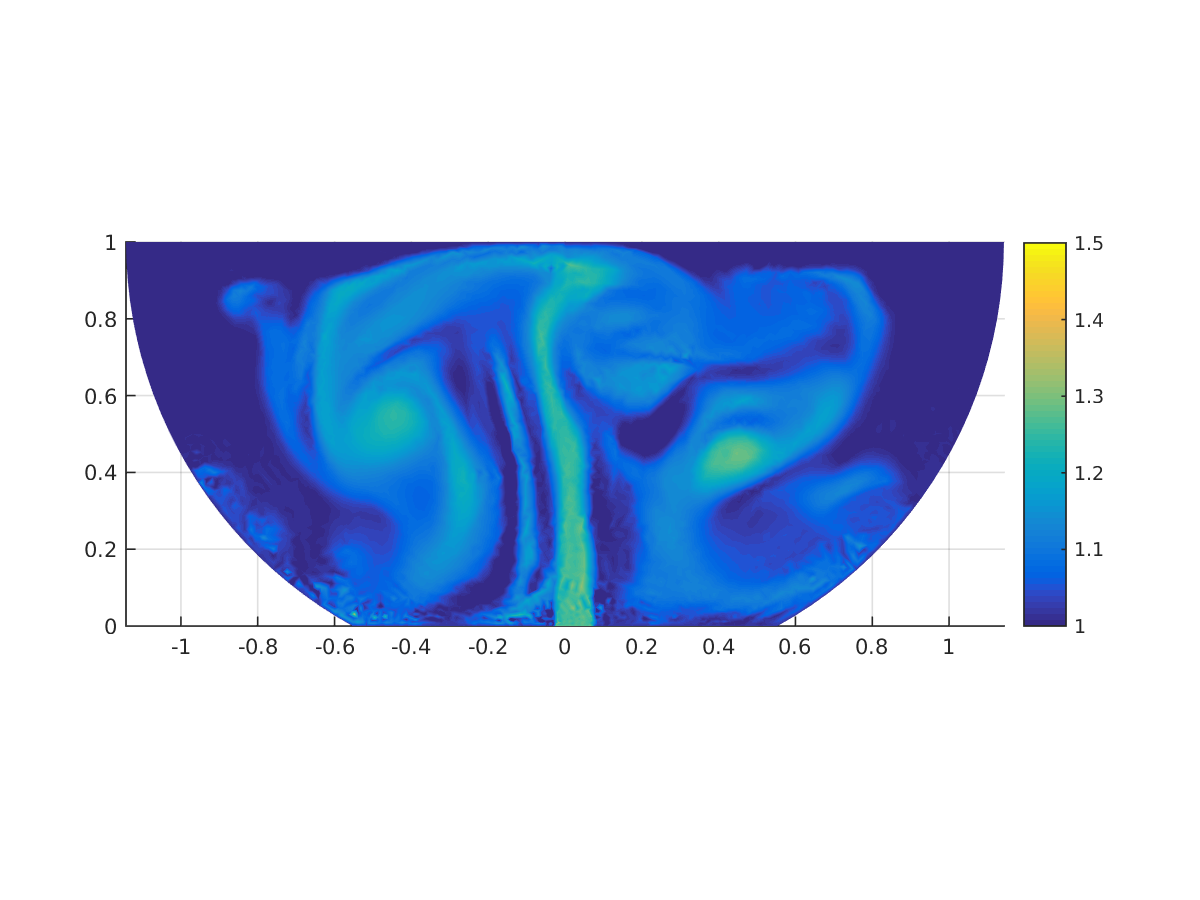}
    \caption{$T_s = 1.350$.}
  \end{subfigure} \begin{subfigure}[b]{0.32\textwidth}
    \includegraphics[width=\textwidth,trim = 0mm 35mm 0mm 35mm,clip]{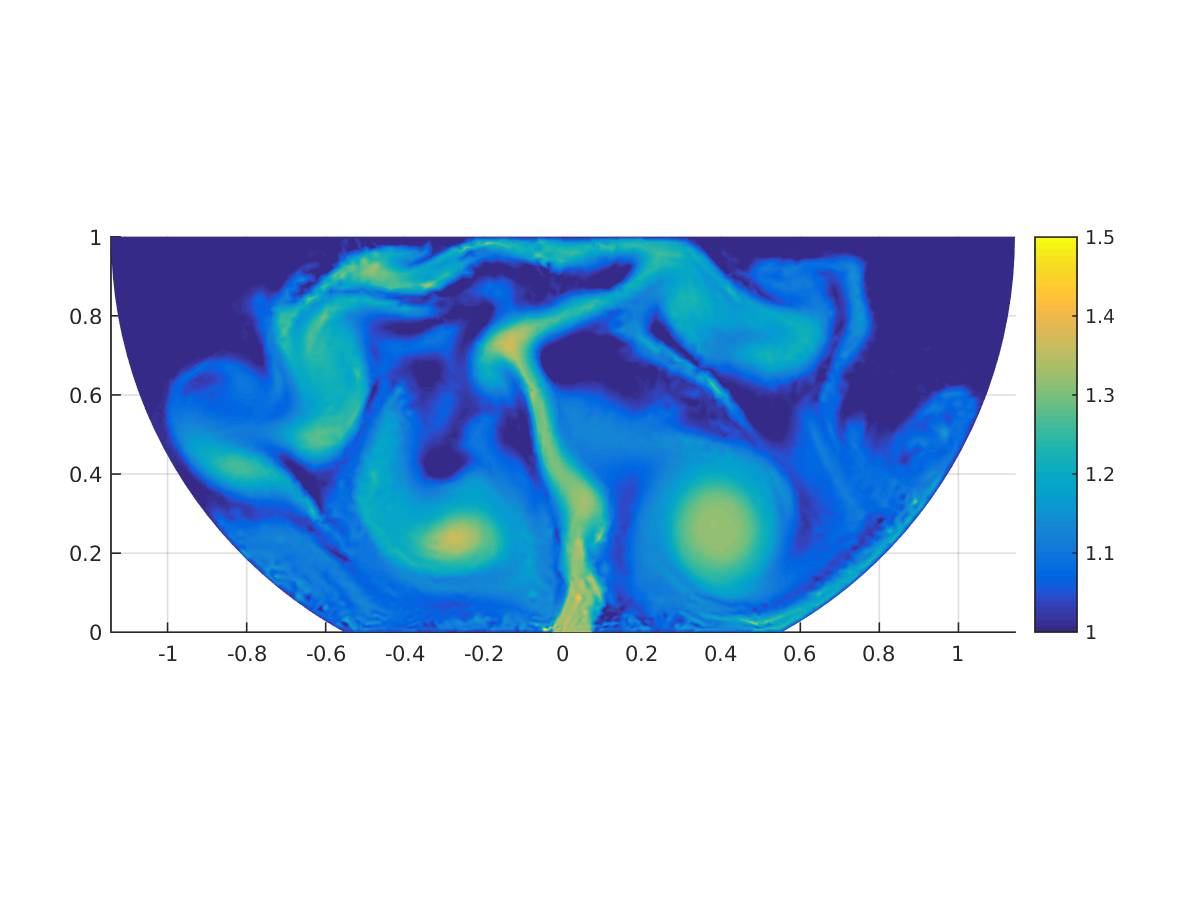}
    \caption{$T_s = 1.431$.}
  \end{subfigure} \begin{subfigure}[b]{0.32\textwidth}
    \includegraphics[width=\textwidth,trim = 0mm 35mm 0mm 35mm,clip]{{fire1.4859x0.0236}.png}
    \caption{$T_s = 1.486$.}
  \end{subfigure}
  \caption{Temperature fields at $x_s = 0.024m$ for the 5 values of temperature source, time $t=10s$ after startup.}
  \label{fig:fixedloc}
\end{figure}

\begin{figure}
  \centering
  \begin{subfigure}[b]{0.32\textwidth}
    \includegraphics[width=\textwidth,trim = 0mm 35mm 0mm 35mm,clip]{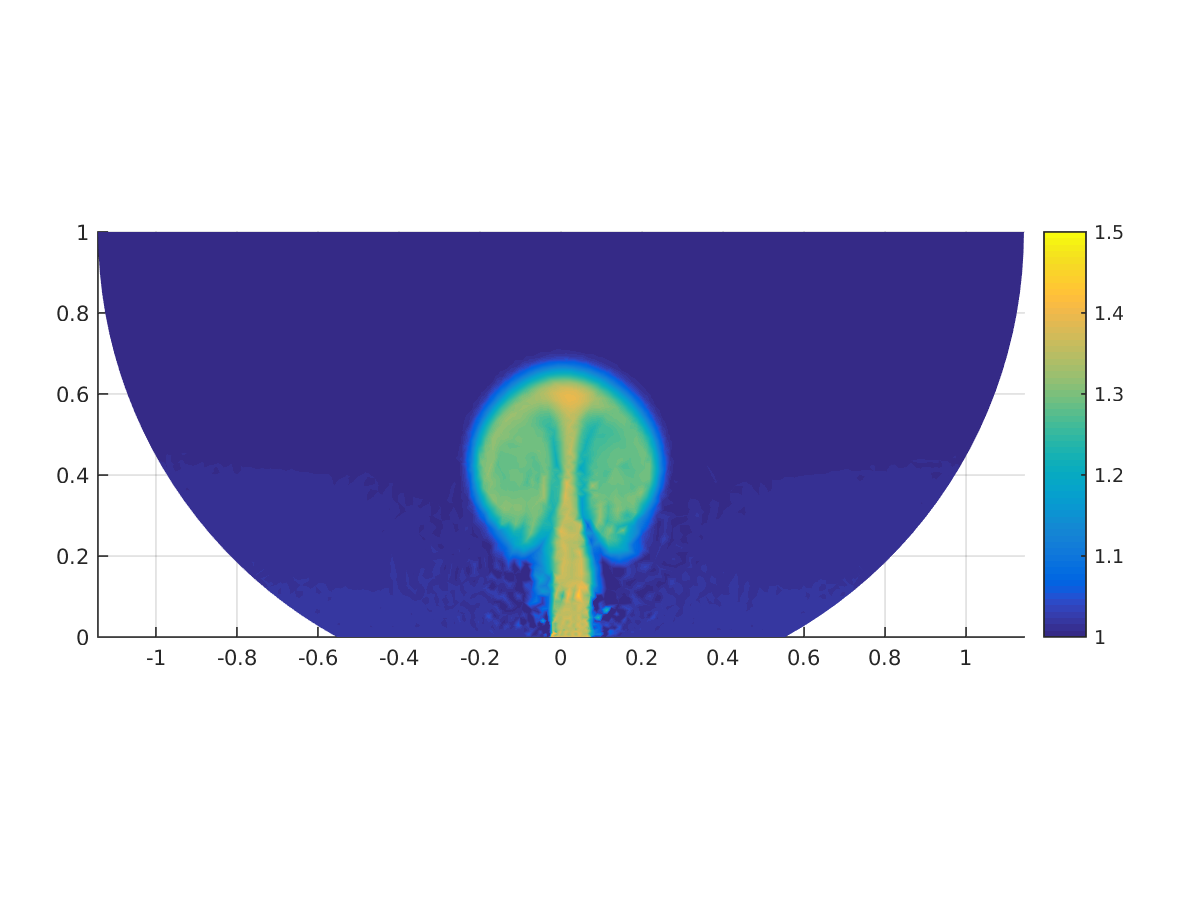}
    \caption{$t = 2s$.}
  \end{subfigure} \begin{subfigure}[b]{0.32\textwidth}
    \includegraphics[width=\textwidth,trim = 0mm 35mm 0mm 35mm,clip]{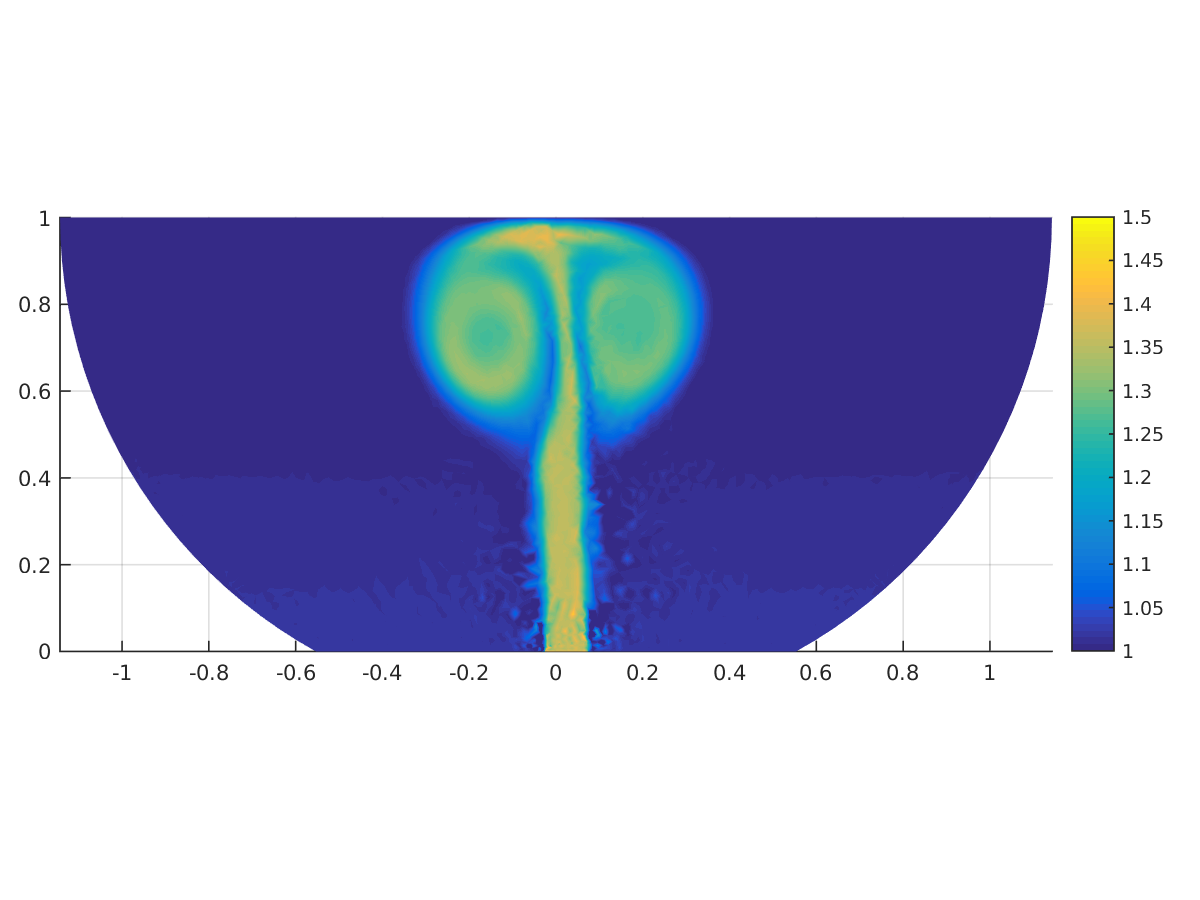}
    \caption{$t = 3s$.}
  \end{subfigure} \begin{subfigure}[b]{0.32\textwidth}
    \includegraphics[width=\textwidth,trim = 0mm 35mm 0mm 35mm,clip]{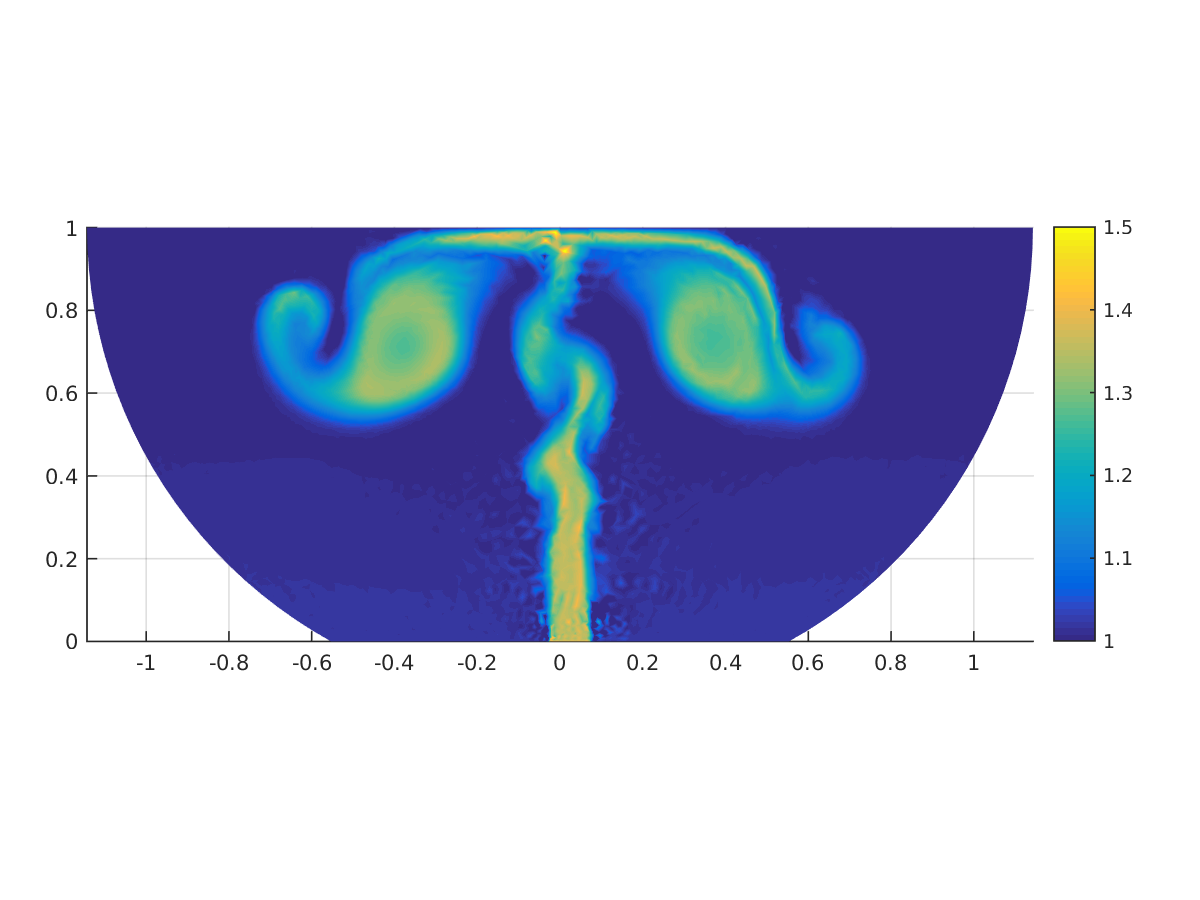}
    \caption{$t = 4s$.}
  \end{subfigure} \begin{subfigure}[b]{0.32\textwidth}
    \includegraphics[width=\textwidth,trim = 0mm 35mm 0mm 35mm,clip]{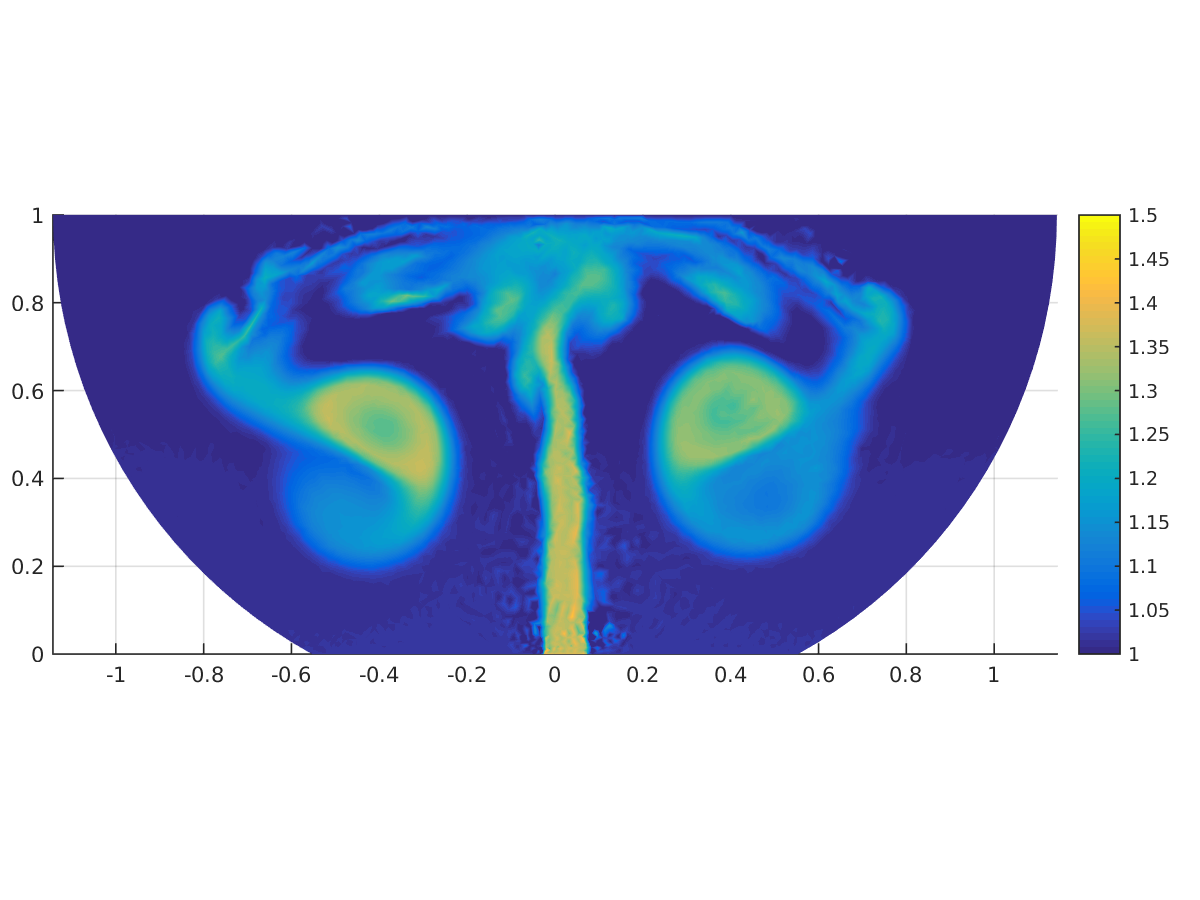}
    \caption{$t = 5s$.}
  \end{subfigure} \begin{subfigure}[b]{0.32\textwidth}
    \includegraphics[width=\textwidth,trim = 0mm 35mm 0mm 35mm,clip]{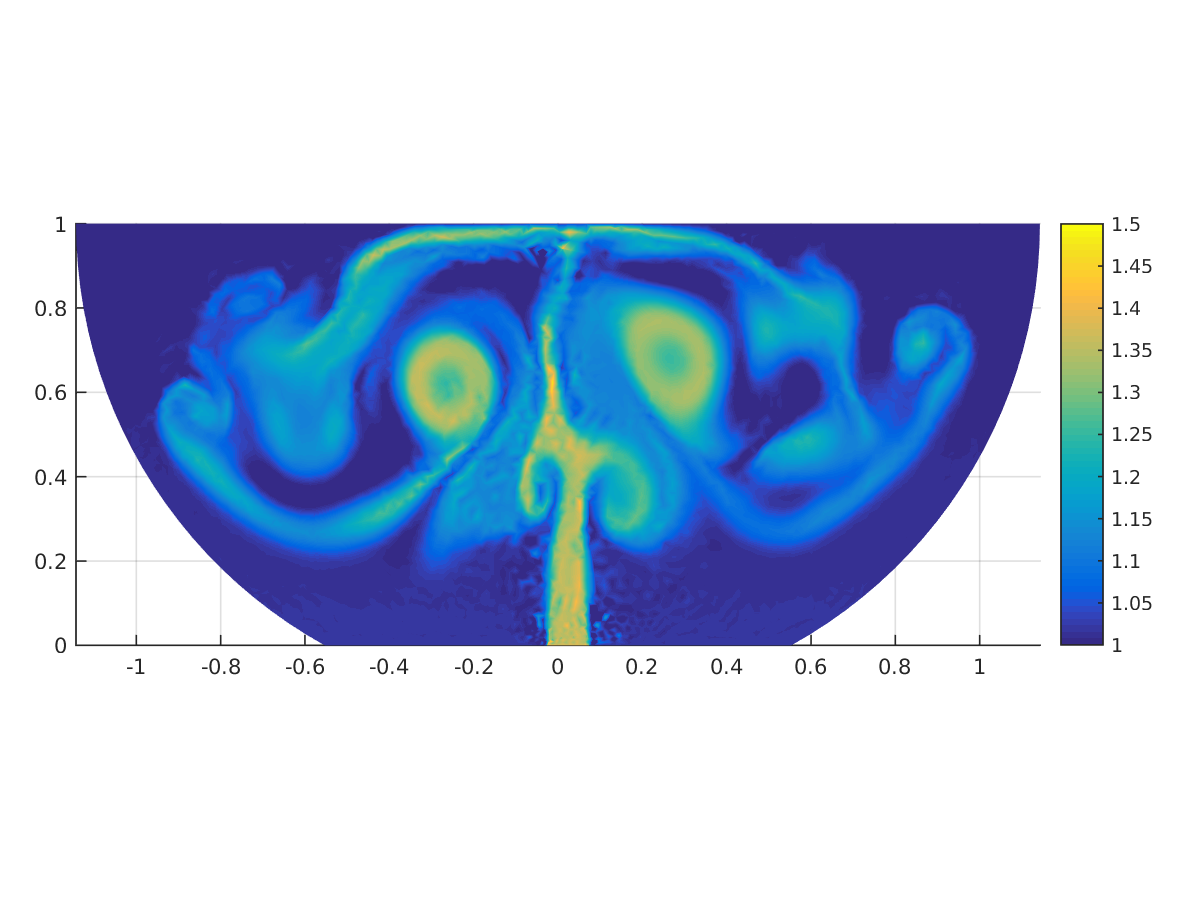}
    \caption{$t = 6s$.}
  \end{subfigure} \begin{subfigure}[b]{0.32\textwidth}
    \includegraphics[width=\textwidth,trim = 0mm 35mm 0mm 35mm,clip]{{fire1.4859x0.0236t7}.png}
    \caption{$t = 7s$.}
  \end{subfigure}
  \caption{Temperature field time evolution for $T_s = 1.486$, $x_s = 0.024$ case.}
  \label{fig:timedependent}
\end{figure}

Sample flowfield snapshots of temperature are displayed in
Figures~\ref{fig:fixedheat},~\ref{fig:fixedloc},
and~\ref{fig:timedependent}. These figures illustrate the wide range
of spatio-temporal flow behaviors that are possible when the fire
source location and temperature are varied, and motivates a study
aimed at quantifying the statistics of some measure of the flow given
parameterized uncertainty in the fire source location and temperature.

\clearpage
\section{Uncertainty Quantification using Polynomial Chaos Expansions (PCE)}
\label{sec:uqmethod}

Motivated by the wide range of possible flow patterns displayed in
Figures~\ref{fig:fixedheat},~\ref{fig:fixedloc},
and~\ref{fig:timedependent}, our goal is to vary the fire source
location $x_S$ and temperature $T_S$ at the base of the cargo hold and
quantify how this affects the flow dynamics, while using a minimal
number of DG simulations. We wish to construct a surrogate model that
explicitly describes this input-output behavior, which can be
inexpensively sampled and analyzed for statistical correlations and
trends. As a result of these requirements, we choose to use Polynomial
Chaos Expansions (PCE) as our framework for performing UQ.

We give a brief overview of this approach below; further details can
be found in introductory references on PCE
methods\cite{ghanem_book,xiu_book,lemaitre}.

Let~$Z=(Z_1,Z_2)$ be a vector of standard random variables that
parameterizes the uncertainty in the cargo hold fire simulations. Let
$d=2$ be the number of random parameters. We are interested in the
corresponding uncertainty of some measure of the fluid dynamics,
represented by $y(Z)$.  In our setting, $Z_1$ is a coordinate that
parameterizes the fire source location along the base of the cargo
hold, and $Z_2$ is a coordinate that parameterizes the fire source
temperature. The output, $y(Z)$, is some measure of the fire dynamics;
for example, the temperature at specific locations along the ceiling
of the cargo hold, at different sample times.

The goal of the method is to represent $y(Z)$ in terms of some basis
functions~$\Phi_i$. Assuming (for ease of exposition) that $y(Z)$ is
scalar-valued, we write:
\begin{equation}
  \label{eq:1}
  y(Z) = \sum_{|i|=0}^N y_i \Phi_i(Z).
\end{equation}
Here, $i=(i_1,i_2)$ is a multi-index, and $|i|=i_1 + i_2$.  We
define an inner product on the space of functions of the random
variables by
\begin{equation}
  \label{eq:2}
  \ip<f,g> = \int_\Gamma f(Z) g(Z) \rho(Z)\,dZ,
\end{equation}
where $\rho(Z)$ denotes the probability density function of $Z$, and
has support $\Gamma$.  A fundamental insight in PCE methods is to
employ basis functions that are orthonormal with respect to this
inner product, so that
\begin{equation}
  \label{eq:3}
  \ip<\Phi_i,\Phi_j> = \delta_{ij},
\end{equation}
where $\delta_{ij}=1$ if $i=j$, and $0$ if $i\ne j$. In particular,
a multivariate basis polynomial $\Phi_i$ may be written as
\begin{equation}
\Phi_i(Z) = \prod_{k=1}^d \phi_{i_k}(Z_k),
\end{equation}
where $\phi_n$ is a (univariate) polynomial of degree~$n$. The $\{
\phi_n\}$ will be a basis of orthogonal polynomials chosen so that the
orthogonality condition~\eqref{eq:3} is satisfied. For example, if we
work with uniformly distributed random variables, then our basis
polynomials would be the multivariate Legendre polynomials.

The coefficients $y_i$ in the expansion~\eqref{eq:1} may be determined
by taking an inner product with $\Phi_j$: because the $\Phi_j$ are
orthonormal, we have
\begin{equation}
  \label{eq:4}
  y_j = \ip<y,\Phi_j>.
\end{equation}
Note that one could also take $y(Z)$ to be a vector of several different
aerodynamic quantities of interest: in this case, the coefficients~$y_i$ in the
expansion~\eqref{eq:1} are vectors, and each component of $y_i$ is determined by
an equation such as~\eqref{eq:4}, for the corresponding component of~$y$.

The important issue now is how we choose to approximate the projection
integrals in~\eqref{eq:4}. A possible choice is to use Gauss
quadrature, in which the function $y(Z)$ is evaluated on a grid
consisting of the tensor product of $d$ separate 1-D quadrature point
sets in parameter space. A drawback of this method is that it suffers from
the curse of dimensionality; however, since our uncertain parameter
space is only two-dimensional, we choose to use this method.

After computing the PCE surrogate, statistical quantities of interest
readily follow. The probability density function of the output
quantities may be approximated by Monte Carlo sampling of the PC
expansion. The mean $\mu$ and variance $\sigma^2$ of the PC model may
be analytically computed as a function of the model coefficients:
\begin{equation}
  \label{eq:MeanVar}
  \begin{aligned}
    \mu &= y_0 \\
    \sigma^2 &= \sum_{|i|=1}^N y_i^2 \|\phi_i\|^2
  \end{aligned}
\end{equation}
Sobol indices, which provide a metric of the relative ``importance''
of each of the uncertain parameters on the output, may also be
analytically computed from the PC model
coefficients~\cite{Sudret2007}. Specifically, the ``total'' Sobol
index $T_i$ is defined as the fraction of the total variance
contributed by all those polynomials in the PC expansion which involve
$Z_i$:
\begin{equation}
  \label{eq:Sobol}
  \begin{aligned}
    T_i &= \frac{\mathbb{E} [ \var(y|Z_{-i}) ]}{\var(y)},\qquad i = 1,\ldots,d \\
    &= \frac{1}{\sigma^2} \sum_{j} y_j^2 \|\phi_j\|^2
  \end{aligned}
\end{equation}
where $j$ in the above expression is understood to index only those
terms in the PC expansion which involve parameter $Z_i$, $Z_{-i}$
denotes all parameters except $Z_i$, and $\mathbb{E}[\cdot]$ denotes
the expected value.

\section{Case Study: 2-D Cargo Hold Fire with Uncertain Location/Temperature}
\label{sec:results}

In this section, we apply the tools discussed to study a test
problem in which both the fire source location and temperature are
independent, uncertain parameters with some joint probability
distribution $\rho(Z)$. We choose to equip both parameters with a
uniform distribution. We assume that the range of possible fire source
locations consists of the right half of the cargo hold floor. This is
done in order to study the effect of spatial asymmetry on the UQ
problem. We assume that the range of possible fire source temperatures
is given by the interval $[ 1.2, 1.5 ] \times T_{\infty}$.

Given that both of our parameters are uniformly distributed, our PC
basis consists of the Legendre polynomials. We choose to truncate the
PC expansion~\eqref{eq:1} at total order $N=4$. This implies that we
use a $5\times5$ grid of collocation points in the parameter space to
evaluate the projection integrals~\eqref{eq:4}, corresponding to the
tensor product of the five zeros of the fifth order Legendre
polynomials (suitably shifted/scaled) with themselves. These nodes are
given in Table~\ref{table:simparams}. These are the collocation points
that specify the fire source locations/temperatures that we will
simulate using our DG code.

\begin{table}[h]
  \begin{center}
  \begin{tabular}{|l||l|}
    \hline
    Temperature strength $T_s$, $5 \times 5$ & $1.214,\ 1.269,\ 1.350,\ 1.431,\ 1.486$ \\ \hline
    Temperature location $x_s(m)$, $5 \times 5$ & $0.024,\ 0.116,\ 0.252,\ 0.387,\ 0.480$ \\ \hline 
      \end{tabular}
  \end{center}
  \caption{Discrete simulation parameters for uncertainty quantification study. The parameter sweep is performed using a tensor product of these values.}
  \label{table:simparams}
\end{table}

Quantifying an entire field quantity $u(x,t;Z)$ using PCE is
difficult. This is because the spatio-temporal behavior of the flow
can vary significantly with fire source location and temperature,
which makes it difficult to interpolate in parameter space accurately
using $4^{th}$ order polynomials. Therefore, we focus on a set of
observables more amenable to our techniques, corresponding to the
temperature along a 1-D segment near the cargo hold ceiling. This
observable vector is highly relevant from an engineering standpoint,
since it informs the choice of fire sensor placement.

We denote the temperature along the line segment at height $y = 0.95$
in the cargo hold as $T_C(x,t;Z)$. Intuitively, one would expect a
certain characteristic rise time $t_R(Z)$ of the buoyant plume from
the fire source, which should be dominated by the source temperature
(and possibly affected by source location if the plume interacts with
the cargo walls). We define $t_R(Z)$ as the time required from
the start of the fire to detection at any point on the ceiling. For
early fire detection, we are interested in the ceiling temperature
distribution averaged over a short period of time beginning at
$t_R(Z)$ (for some choice of $Z$). Therefore, we define the
time-averaged ceiling temperature distribution:
\begin{equation}
  \label{eq:distTimeAvg}
  \overline{T_C}(x;Z) = \frac{1}{\Delta t} \int_{t_R(Z)}^{t_R(Z)+\Delta t} T_C(x,t;Z) \; dt \quad ,
\end{equation}

and we quantify uncertainty in the observables $\overline{T_C}(x;Z)$ and 
$t_R(Z)$. We use an averaging time period of $\Delta t = 1 s$. Note
that $t_R(Z)$ is a scalar quantity and hence has one PC expansion
associated with it, whereas $\overline{T_C}(x;Z)$ is a function in $x$
(which is discretized as a vector at discrete locations) and hence has
one PC expansion for each location in $x$ we choose to measure. The
units of $t_R(Z)$ will be seconds; as noted previously,
$\overline{T_C}(x;Z)$ is temperature normalized by the initial bulk
temperature $T_{\infty}$.

\begin{figure}[h]
  \centering
  \begin{subfigure}[t]{0.45\textwidth}
    \includegraphics[width=1\textwidth,trim = 40mm 0mm 40mm 0mm,clip]{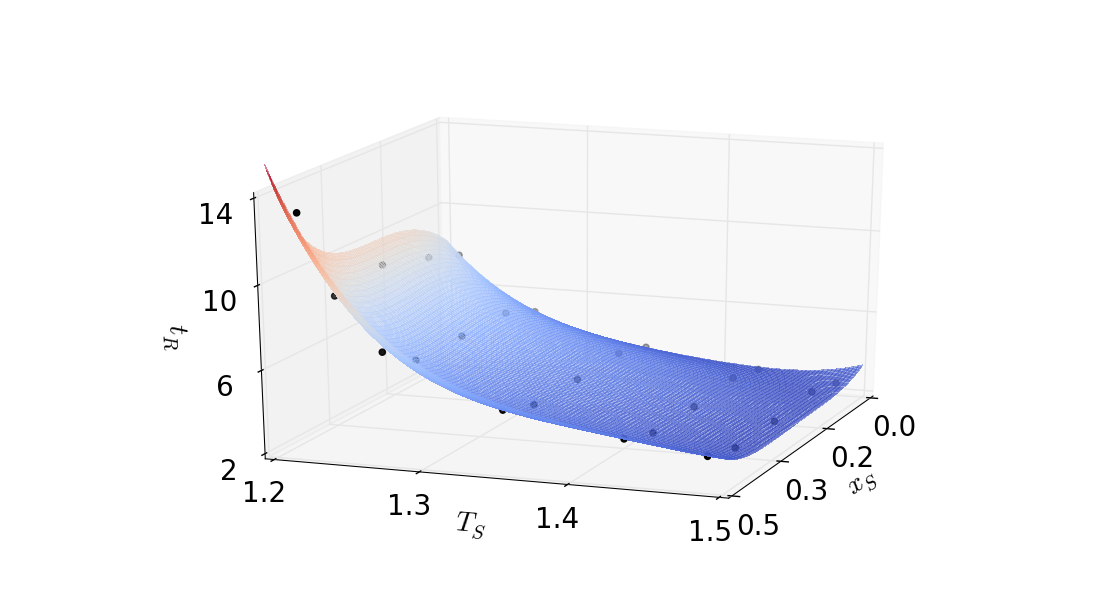}
    \caption{PCE surrogate map of $t_R(Z)$, together with the values at the 25 quadrature nodes.}
  \end{subfigure}
  \begin{subfigure}[t]{0.45\textwidth}
    \includegraphics[width=1\textwidth]{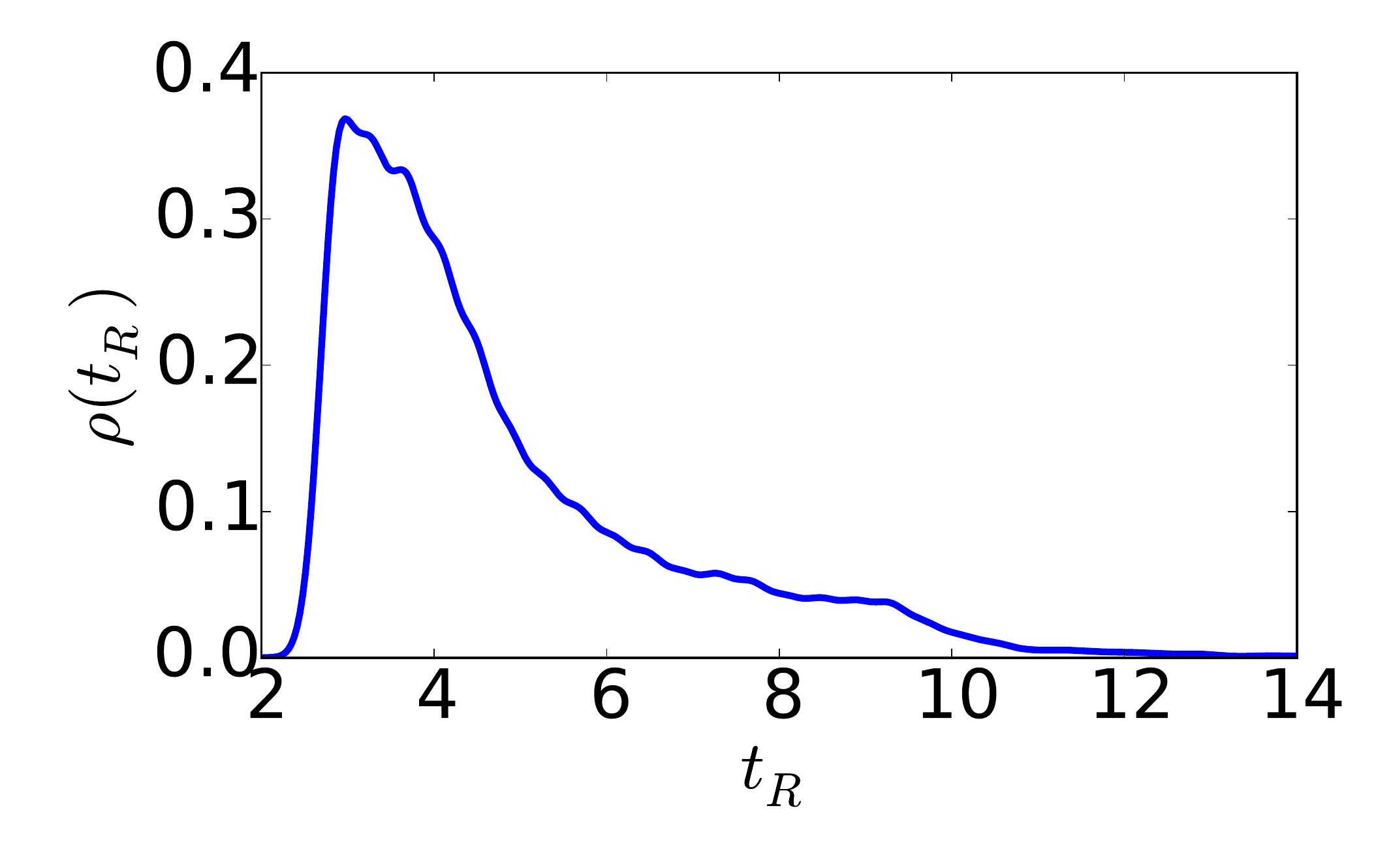}
    \caption{Probability density function $\rho(t_R(Z))$ (approximated using 10,000 random samples of the PCE surrogate).}
  \end{subfigure}
  \caption{PCE surrogate for $t_R(Z)$.}
  \label{fig:RiseTime}
\end{figure}

\begin{figure}[h]
  \centering
  \begin{subfigure}[b]{0.32\textwidth}
    \includegraphics[width=\textwidth,trim = 0mm 35mm 0mm 25mm,clip]{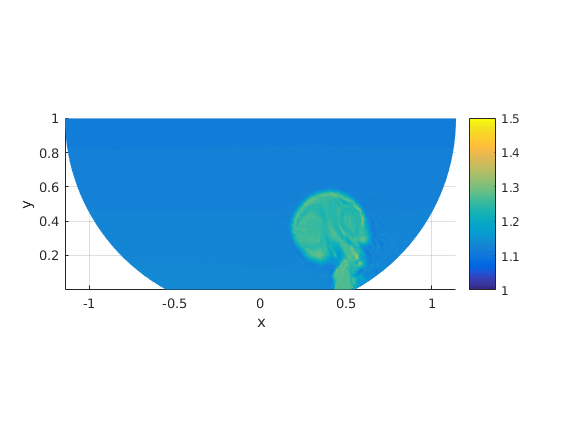}
    \caption{$t = 6s$.}
  \end{subfigure} \begin{subfigure}[b]{0.32\textwidth}
    \includegraphics[width=\textwidth,trim = 0mm 35mm 0mm 25mm,clip]{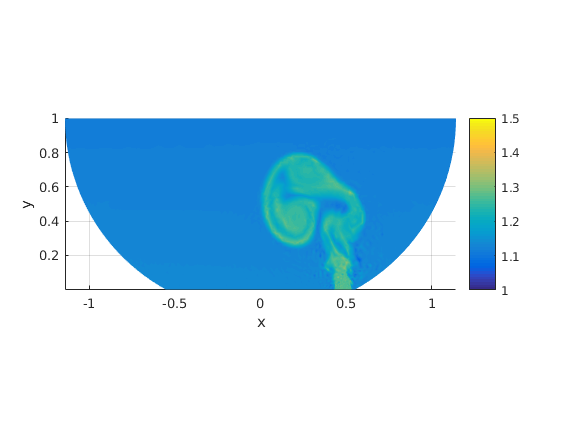}
    \caption{$t = 8s$.}
  \end{subfigure} \begin{subfigure}[b]{0.32\textwidth}
    \includegraphics[width=\textwidth,trim = 0mm 35mm 0mm 25mm,clip]{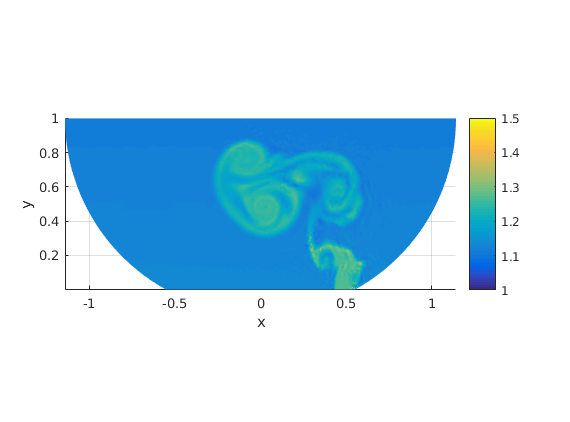}
    \caption{$t = 10s$.}
  \end{subfigure}
  \caption{Temperature field snapshots with $Z = (0.48,1.21)$. The initial plume falls toward the floor without ever touching the ceiling, explaining the unusually long rise time.}
  \label{fig:PlumeFallingOver}
\end{figure}

We first examine the rise time $t_R(Z)$. The PCE surrogate model for
rise times $t_R(Z)$ is shown in Figure~\ref{fig:RiseTime}, along with
statistical quantities in Table~\ref{table:StatsTR}. Examination of
Figure~\ref{fig:RiseTime} and Table~\ref{table:StatsTR} confirms our
hypothesis that the characteristic rise time $t_R$ is dominated by the
source temperature. As shown, source temperature has a Sobol index of
0.95, which means that $95\%$ of the variance in the distribution of
$t_R$ can be attributed to source temperature (either acting alone or
interacting with location). The only portion of parameter space that
really is affected strongly by source location appears to be the
``corner'' area of parameter space where the source is very close to
the cargo hold wall and the source temperature is very low. The result
of this combination of variables is that the initial buoyant plume
``rolls over'' toward the center of the cargo hold and falls back
downward toward the floor before reaching a height of $y=0.95$ (where
we are observing ceiling temperature). This time-dependent behavior is
illustrated in Figure~\ref{fig:PlumeFallingOver}. It is not until
several seconds after this has occured that subsequent buoyant plumes
finally touch the ceiling. This combination of low temperature with a
wall effect is what accounts for the tail of the distribution of
$t_R$.

\begin{table}[h]
  \begin{center}
  \begin{tabular}{|l||l|}
    \hline
    Mean & 4.8 \\ \hline
    Variance & 3.9 \\ \hline
    Sobol Index 1 (Location) & 0.08 \\ \hline
    Sobol Index 2 (Temperature) & 0.95 \\ \hline \hline
      \end{tabular}
  \end{center}
  \caption{Statistical quantities of interest for $t_R(Z)$.}
  \vspace*{-1cm}
  \label{table:StatsTR}
\end{table}

\begin{figure}[h]
  \centering
  \begin{subfigure}[t]{0.45\textwidth}
    \centering
    \includegraphics[width=0.75\textwidth]{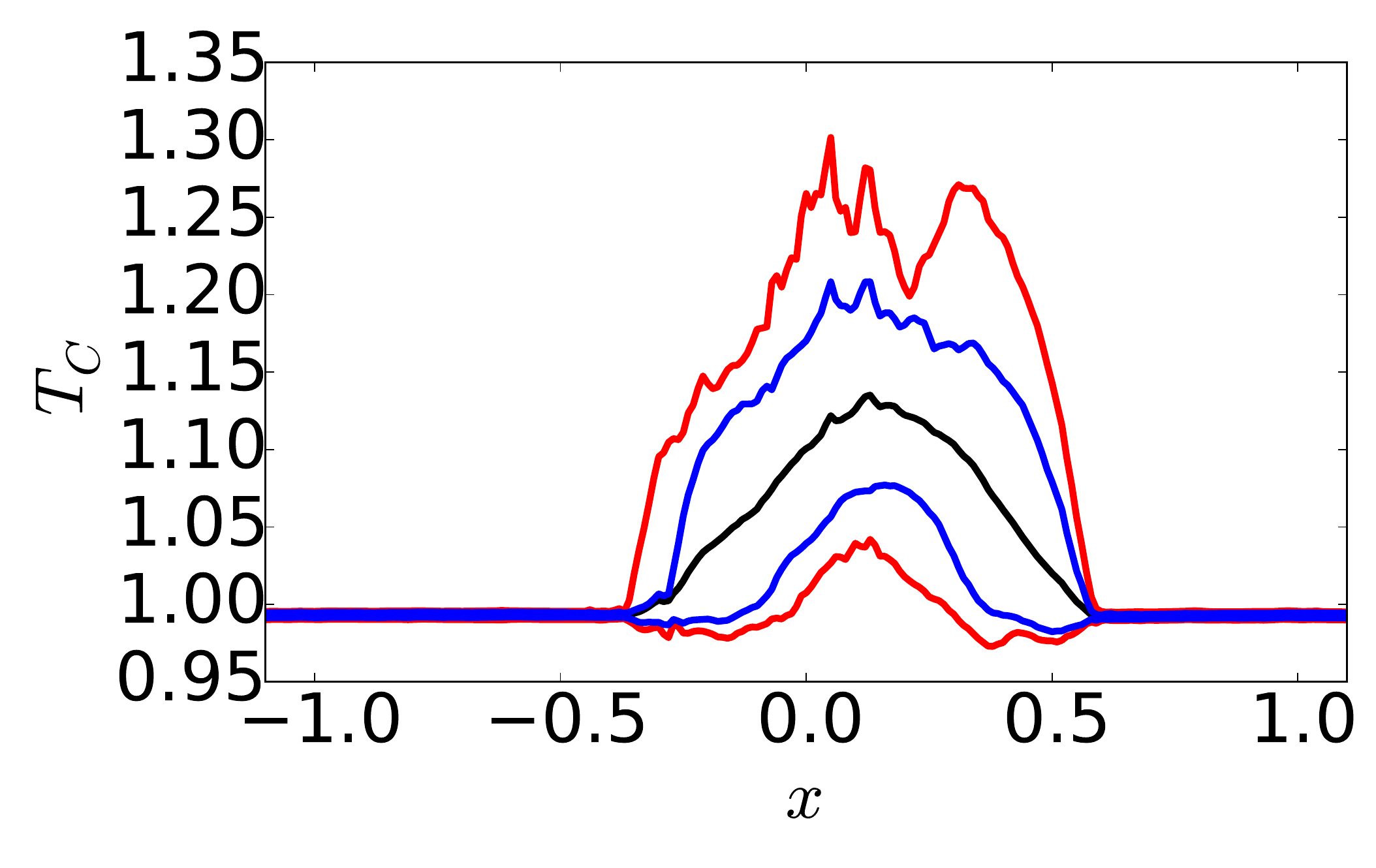}
    \caption{{\it Black}: mean time-averaged ceiling temperature profile. {\it Blue}: 68$\%$ confidence interval. {\it Red}: 95$\%$ confidence interval.}
    \label{fig:CeilingMeanPct}
  \end{subfigure}
  \begin{subfigure}[t]{0.45\textwidth}
    \centering
    \includegraphics[width=0.75\textwidth]{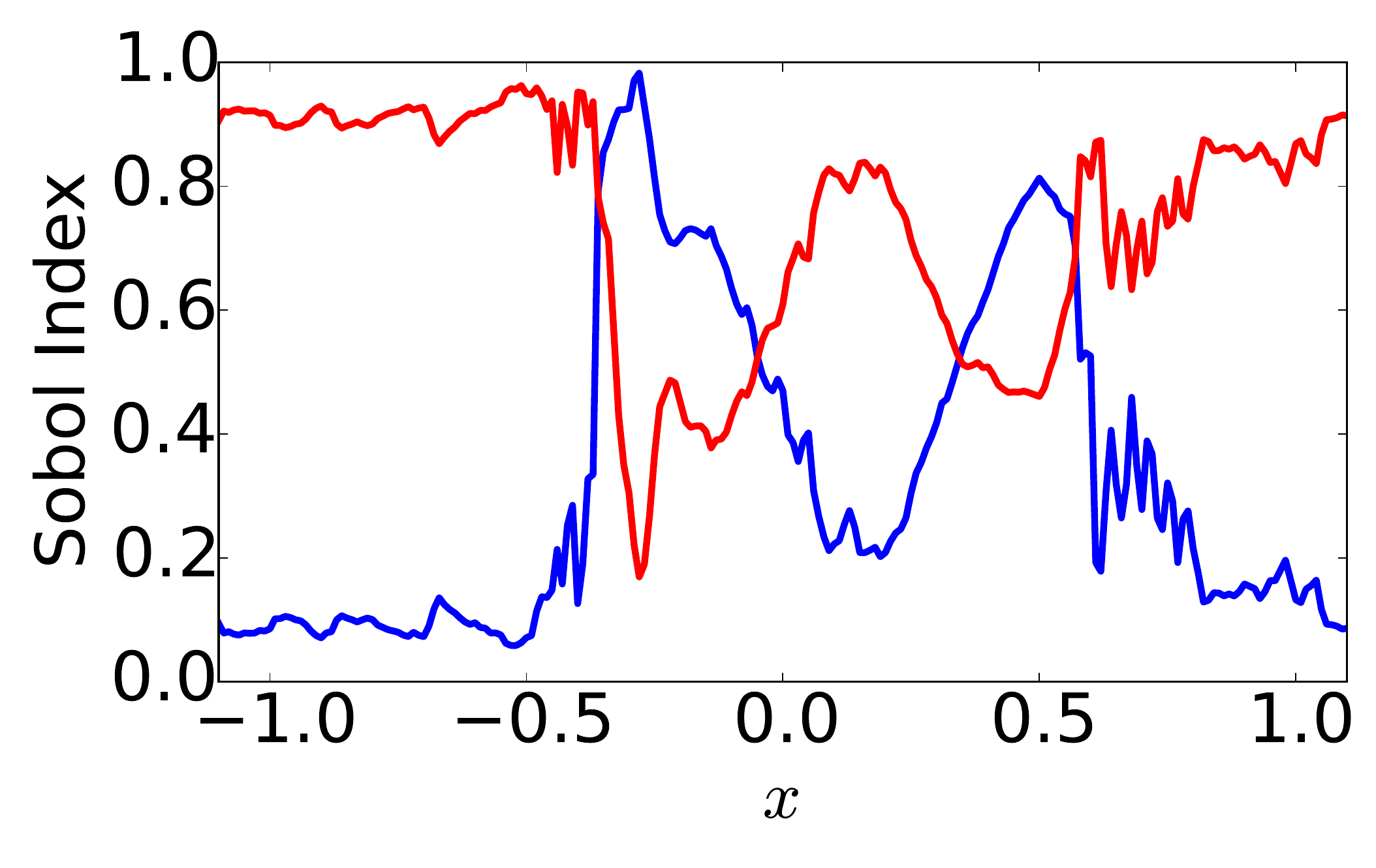}
    \caption{Sobol indices for ceiling temperature at points along the ceiling. {\it Blue}: source location. {\it Red}: source temperature.}
    \label{fig:SobolCeiling}
  \end{subfigure}
  \caption{Statistical quantities of interest for time-averaged ceiling temperature.}
  \label{fig:CeilingMeanAndSobol}
\end{figure}

\begin{figure}[h!]
  \centering
  \includegraphics[width=0.8\textwidth,trim = 40mm 0mm 35mm 0mm, clip]{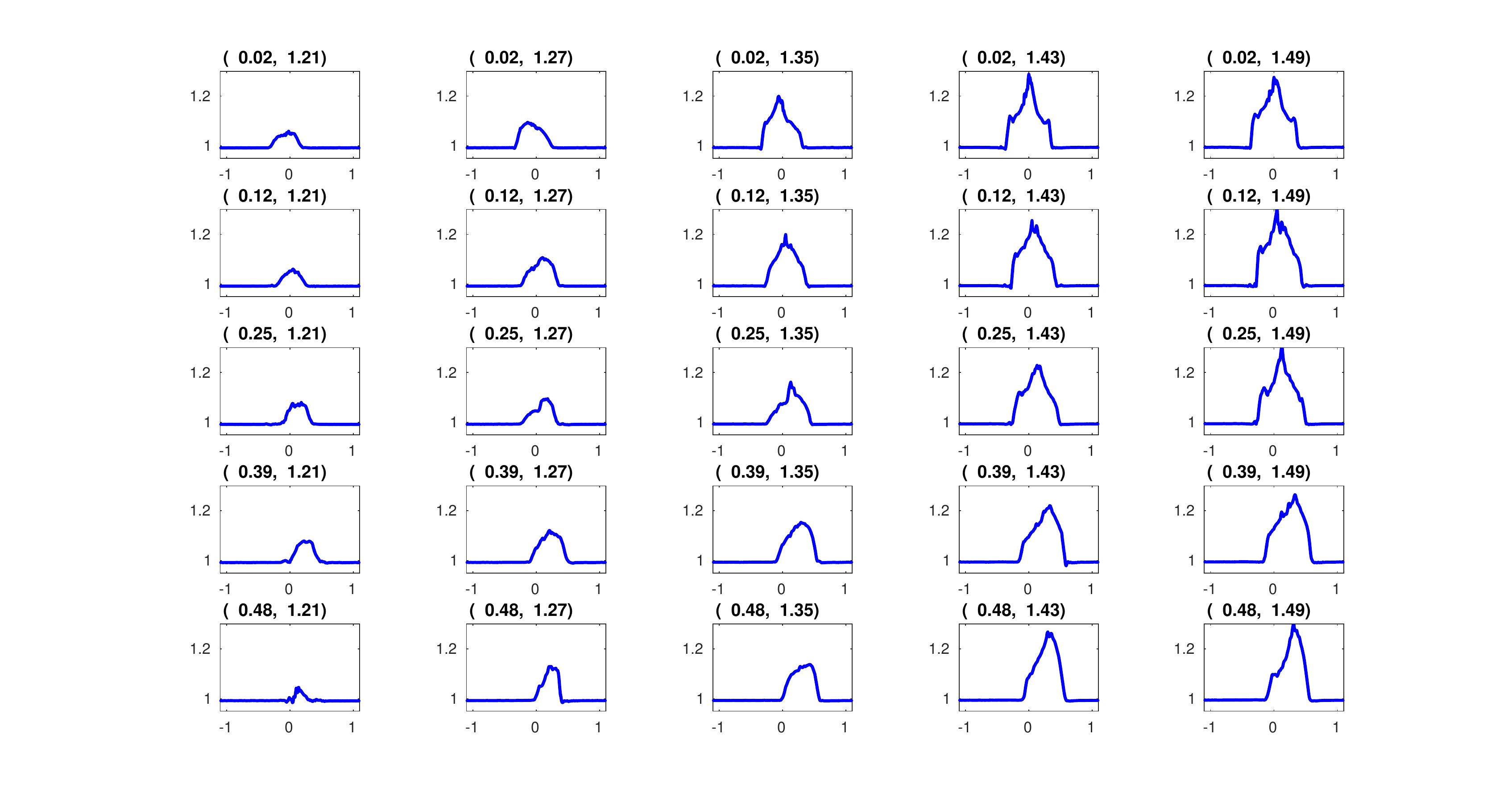}
  \caption{Time-averaged ceiling temperature distributions collected at the 25 quadrature nodes. Each subtitle corresponds to the parameter pair $(x_S, T_S)$.}
  \label{fig:CeilingTempNodes}
\end{figure}

\begin{figure}[h!]
  \centering
  \includegraphics[width=0.8\textwidth,trim = 40mm 0mm 35mm 0mm,clip]{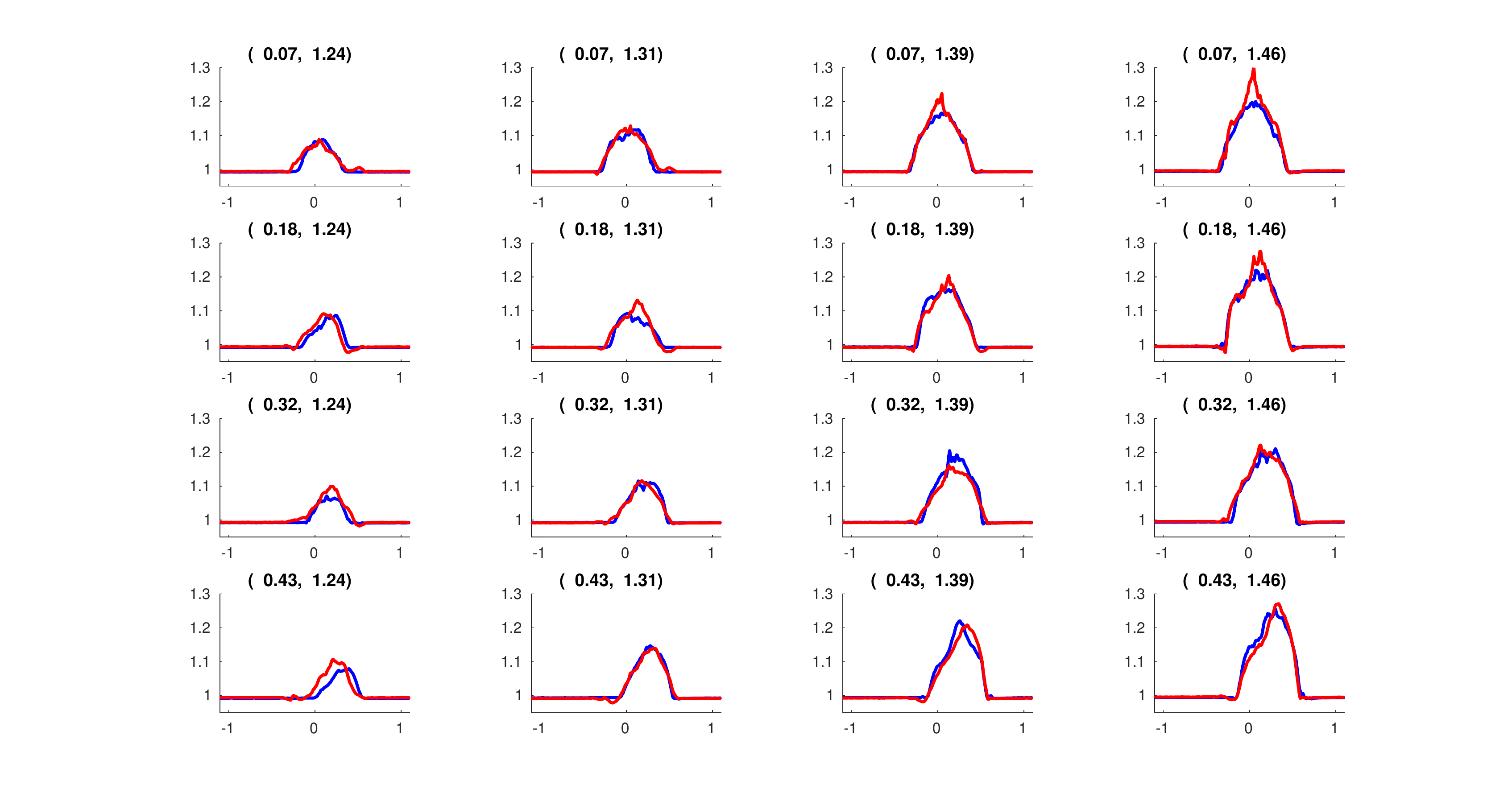}
  \caption{Time-averaged ceiling temperature distributions collected at points on the 4x4 mesh which is dual to the 5x5 mesh. Data are displayed in blue; PC models are displayed in red. Each subtitle corresponds to the parameter pair $(x_S,T_S)$.}
  \label{fig:PCEVerificationCeiling}
\end{figure}

We next turn our attention to the time-averaged ceiling temperature
distribution $\overline{T_C}(x;Z)$. The time-averaged ceiling
temperature distributions at the quadrature nodes are shown in
Figure~\ref{fig:CeilingTempNodes}. The mean distribution along with
confidence intervals -- computed from Monte Carlo samples of the PCE
surrogate -- is shown in Figure~\ref{fig:CeilingMeanPct}. The accuracy
of the PC model for ceiling temperature can be verified by comparing
the PC interpolation to data at various points in parameter
space. This is done in Figure~\ref{fig:PCEVerificationCeiling}, which
confirms that our the PC model provides reasonably accurate
interpolation.

We can also examine the total Sobol indices as a function of $x$ for
the ceiling temperature observable, which indicate which of the two
uncertain parameters best explains the variance in the ceiling
temperature. These Sobol indices are displayed in
Figure~\ref{fig:SobolCeiling}. As can be seen, source temperature is
the dominant parameter in the area around the maximum of the mean
profile. The peripheral areas are dominated by source location. The
explanation of this phenomenon is natural: source temperature controls
the intensity of the temperature fluctuations observed on the ceiling
where they are hottest, but source location determines whether or not
temperature fluctuations are actually observed at all in the
peripheral areas.

\begin{figure}[h]
  \centering
  \begin{subfigure}[t]{0.45\textwidth}
    \centering
    \includegraphics[width=0.75\textwidth]{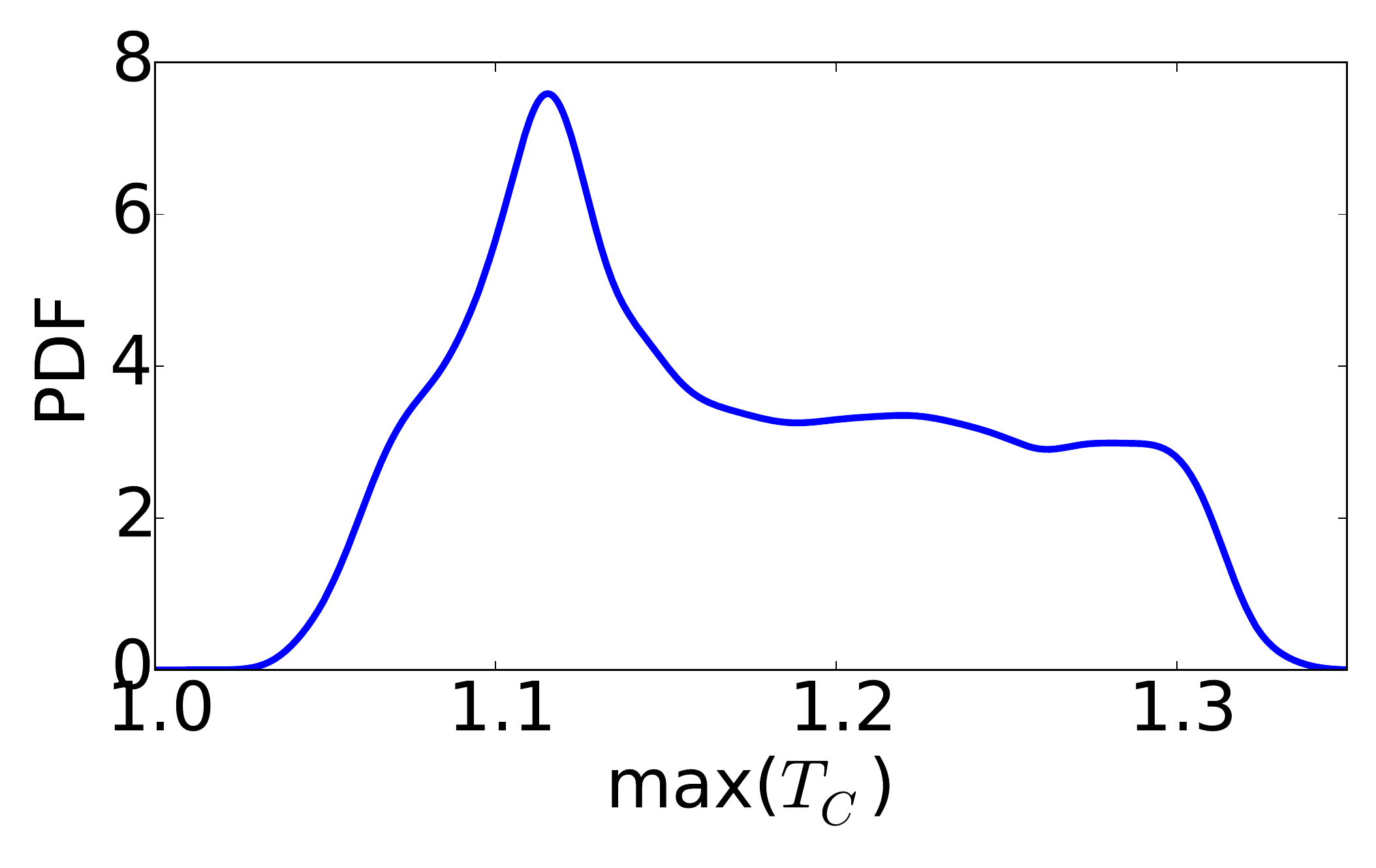}
    \caption{Distribution of the maximum value of the time-averaged ceiling temperature.}
    \label{fig:maxCeilingTempDistribution}
  \end{subfigure}
  \begin{subfigure}[t]{0.45\textwidth}
    \centering
    \includegraphics[width=0.75\textwidth]{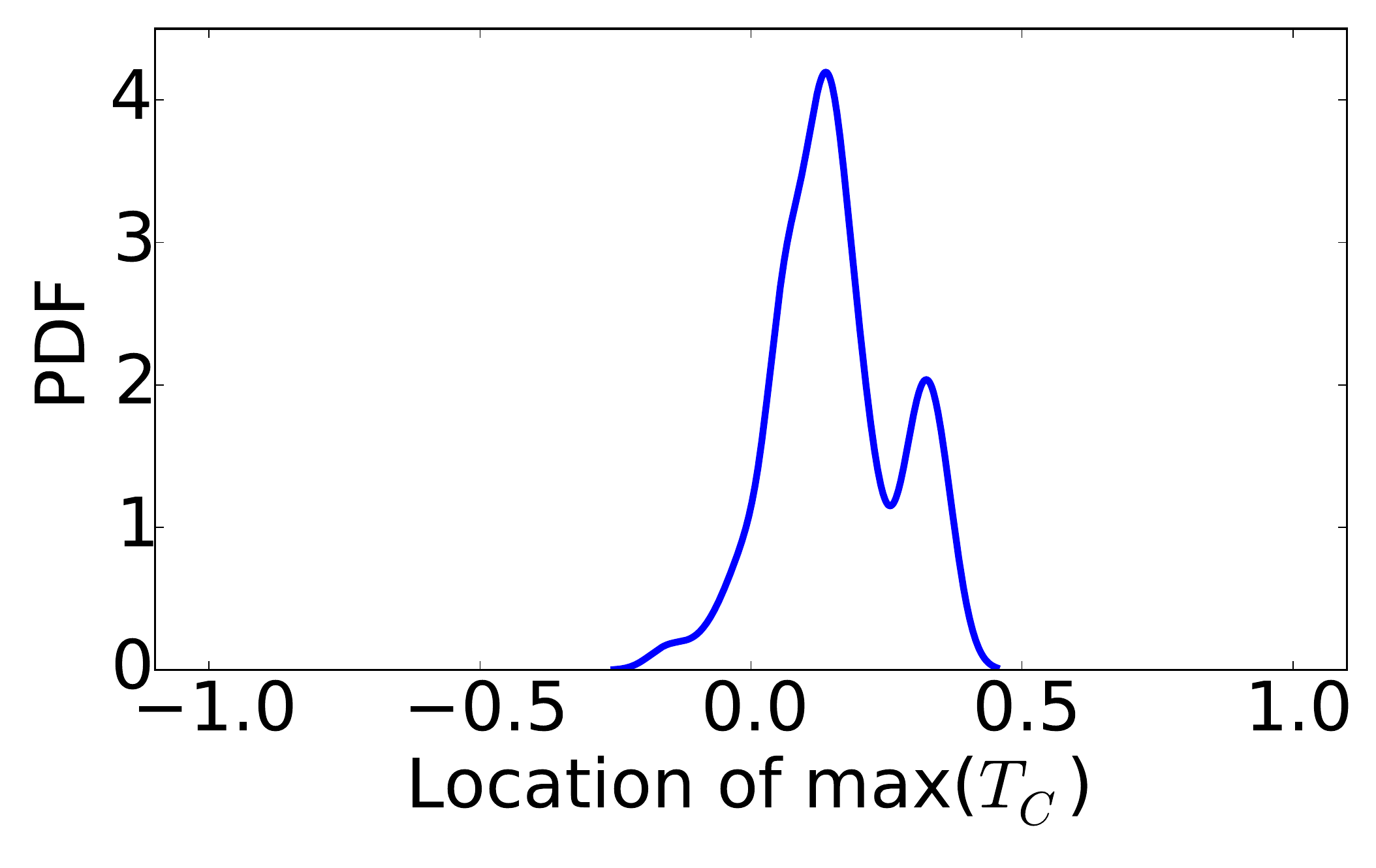}
    \caption{Distribution of the location of the maximum value of the time-averaged ceiling temperature.}
    \label{fig:maxCeilingTempLocationDistribution}
  \end{subfigure}
  \caption{Distributions of maximum ceiling temperature value and location. Computed from 10,000 Monte Carlo samples of the PCE surrogate.}
  \label{fig:maxCeilingTempAndLocation}

\end{figure}

Having a PCE surrogate for ceiling temperature also means that we can
compute the statistics of any quantity derived from it. Two
particularly relevant examples of this include the maximum value of
$\overline{T_C}(x;Z)$ as well as its location along the ceiling. We
display these statistics in
Figure~\ref{fig:maxCeilingTempAndLocation}. We see that a wide range
of maximum ceiling temperatures are possible, with a skew toward lower
maximum values. We also see a clear skew in the location of the
maximum ceiling temperature to the right of the center (as would be
expected from the asymmetry in the source location). Computing
correlations between these output quantities and our uncertain
parameters confirms what one would expect -- source temperature
dominates the maximum value of the ceiling temperature, whereas source
location dominates its location.

Combining all of this information together gives a clear and
insightful view of the physics of our cargo hold problem. We see that
the main effects of increasing temperature are to increase the maximum
ceiling temperature, and to decrease the rise time. The main effect of
location is to influence whether or not fluctuations in ceiling
temperature are observed in the peripheral regions of the ceiling. The
fact that we observe these intuitive trends in our surrogate model
gives us further validation of the claim that PCE methods may provide
a method for UQ which is not only efficient, but also {\it accurate}
for this class of problems.

 Of course, the main usage of these UQ tools is not just to confirm
 intuition, but to {\it quantify} it. We see that, on average, we can
 expect a ceiling temperature distribution which is roughly symmetric
 between the limits $y \in [-0.35, 0.60]$, with a maximum around $y =
 0.125$. We can also give confidence intervals on the mean ceiling
 temperature distribution (Figure~\ref{fig:CeilingMeanAndSobol}), and
 estimate the probability distributions for the value and location of
 the maximum ceiling temperature
 (Figure~\ref{fig:maxCeilingTempAndLocation}).

\section{Conclusions}

The purpose of this paper was establish a framework for performing
efficient, accurate investigations of the statistical variations in
cargo hold fires that occur due to parameterized uncertainty in the
boundary conditions. We address two related problems -- increasing the
numerical accuracy of the CFD simulation, and uncertainty
quantification. Higher order numerical accuracy is necessary because
traditional finite-volume schemes require a prohibitively fine mesh in
order to resolve the vortex-dominated flows seen in cargo hold fire
solutions. The need for uncertainty quantification stems from the fact
that the boundary conditions of the cargo hold fire will always be
fundamentally unpredictable, since one can never know {\it a priori}
exactly where the fire will start, how hot it will be, how much
luggage clutter there is, etc.

In order to provide greater simulation accuracy, we developed an
in-house discontinuous Galerkin (DG) flow solver for the compressible
Navier-Stokes equations with buoyancy effects. This code also features
an unstructured mesh suitable for complex geometries. To make
uncertainty quantification feasible, we first reduced the problem from
quantifying the full flow field to quantifying {\it measures} of the
flow field -- a characteristic rise time of the buoyant flow, and a
time-averaged ceiling temperature distribution. This made the problem
amenable to treatment with spectral expansion methods, and so we used
PCE as the tool to efficiently and accurately quantify the effects of
fire source location and temperature. A case study of a 2D cargo hold
geometry in which the fire source location and temperature were
uncertain confirmed that PCE tools provide a viable UQ approach, and
keep the number of required CFD simulations to a minimum.

We are currently working to extend these methods to 3D cargo hold fire
configurations. We are also planning to investigate methods for
accounting for geometric uncertainty in cargo hold luggage clutter,
which was not accounted for in our empty cargo hold geometries.

\section{Acknowledgments}

This research was supported under the Federal Aviation Administration
(FAA) Joint University Program (JUP). The authors would also like to
thank Ezgi Oztekin for helpful discussions on previous cargo hold fire
research as well as a tour of the cargo hold fire testing facilities
at the FAA Tech Center.

\bibliographystyle{aiaa}	    
\bibliography{AIAA2016References}   

\end{document}